\documentclass[journal,twoside,web]{ieeecolor}
\usepackage{generic}
\usepackage{cite}
\usepackage{amsmath,amssymb,amsfonts}
\usepackage{algorithmic}
\usepackage{graphicx}
\usepackage{algorithm,algorithmic}
\usepackage{hyperref}
\usepackage{subfigure}
\usepackage{textcomp}
\usepackage{booktabs}
\makeatletter

\newcommand{\Rmnum}[1]{\expandafter\@slowromancap\romannumeral #1@}
\makeatother
\usepackage{epstopdf}
\hypersetup{hidelinks=true}
\usepackage{textcomp}
\def\BibTeX{{\rm B\kern-.05em{\sc i\kern-.025em b}\kern-.08em
    T\kern-.1667em\lower.7ex\hbox{E}\kern-.125emX}}
\begin{document}
\title{Resilient Distributed Control for Uncertain Nonlinear Interconnected Systems under Network Anomaly}
\author{Youqing Wang, \IEEEmembership{Senior Member, IEEE}, Ying Li, \IEEEmembership{Student Member, IEEE}, Thomas Parisini, \IEEEmembership{Fellow, IEEE} and Dong Zhao, \IEEEmembership{Senior Member, IEEE}
\thanks{Youqing Wang and Ying Li is with the College of Information Science and Technology, Beijing University
 of Chemical Technology, Beijing 100029, China (e-mail: wang.youqing@ieee.org; li.ying@ieee.org).}
\thanks{Thomas Parisini is with the Department of Electrical and Electronic Engineering, Imperial College London, London SW7 2AZ, UK, with the Department of Engineering and Architecture, University of Trieste, 34127 Trieste, Italy, and with the KIOS Research and Innovation Center of Excellence, University of Cyprus, CY-1678 Nicosia, Cyprus (email: t.parisini@imperial.ac.uk).}
\thanks{Dong Zhao is with the School of Cyber Science and Technology, Beihang
University, Beijing 100191, China (e-mail: dzhao@buaa.edu.cn).}
}

\maketitle

\begin{abstract}
We address a distributed adaptive control methodology for nonlinear interconnected systems possibly affected by network anomalies. In the framework of adaptive approximation, the distributed controller and parameter estimator are designed by exploiting a back-stepping approach. The stability of the distributed control system under anomalies is analyzed, where both local and neighboring anomaly effects are considered. To quantify the resilience of the interconnected system under the action of network anomalies, we derive bounds on the duration of each anomaly and the resting time between two consecutive anomalies. Specifically, when each anomaly duration is smaller than our designed upper bound, the interconnected system controlled by the distributed approximation-based controller remains asymptotically stable. Moreover, if the resting time between two consecutive anomalies is larger than the proposed bound, then all signals of the control system are guaranteed to be bounded. In the paper, we show that under the action of the proposed distributed adaptive controller, the interconnected system remains stable in the presence of network anomalies, with both the qualitative and quantitative resilient conditions. Extensive simulation results show the effectiveness of our theoretical results.
\end{abstract}


\section{Introduction}\label{e1}
\IEEEPARstart{C}{yber}-physical systems (CPS) are involved in most critical infrastructure systems, such as power grids, aerospace, transportation, and water systems. Usually, these applications have a large-scale nonlinear interconnected structure characterized by high cyber-security requirements, which inspires and motivates significant research efforts \cite{s1}. In this respect, network anomalies may cause severe damages to critical infrastructure operation.

\subsection{A glimpse of the state of the art}

Cyber-attacks are a form of network anomalies, and numerous results have been reported in this area. Cyber-attacks on CPS can be categorized into three main categories: denial of service (DoS) attacks, deception attacks, and disclosure attacks \cite{a1,a2,a3,a4}. Many research results have been put forward on anomaly detection \cite{a5,a6,a7}, controller design \cite{a8,a9,a10}, and stability analysis of CPS under network anomalies \cite{a11,a12}. Just as a notable example, in \cite{a13}, the S-type function is introduced to deal with the control problem of nonlinear systems under deception attacks, and the proposed adaptive control scheme makes the stability error of the system arbitrarily small. As another example, in \cite{a14}, the classical variable structure hybrid intelligent control method is adopted, and the adaptive law is derived to compensate for cyber-attacks of CPS. Indeed, in \cite{a15},
the state feedback controller is designed under DoS attacks, and a stability criterion for the closed-loop system is given.

Unlike centralized systems, in distributed CPS there are physical-level interconnection and cyber-level information exchange among interconnected subsystems. Thus, when any subsystems are subject to network anomalies, the anomalies effect may spread to other subsystems and threaten the operating performance and stability of the entire interconnected system. For example, in \cite{a16}, the issue of output feedback security control under DoS attacks is studied, and an adaptive security controller based on a extreme learning machine is designed for high-order nonlinear interconnected systems to ensure uniformity and ultimate bounding. In \cite{a17}, the improved residence time method is used to derive the switching state estimator, which overcomes the problem that the state variables of the interfered nonlinear interconnected CPS are unavailable under intermittent DoS attacks. In \cite{a18}, the finite-time $H_\infty$ control problem under random cyber-attacks are considered, and the fuzzy observer and controller are obtained by solving the optimization algorithm under the constraint of linear matrix inequalities.

A CPS is usually nonlinear and uncertain, and neural networks have been widely used due to their global approximation capabilities \cite{s2}. In \cite{a19}, the unknown part of the system is modeled by neural networks, the state under DoS attacks is obtained by the designed dynamic gain observer, and the nonstrict-feedback nonlinear interconnected systems stability analysis is carried out. In \cite{a20}, neural networks are used for the intelligent estimator and the controller designed under the proposed adaptive control architecture to guarantee the ultimate boundedness of the closed-loop system when the actuator is attacked by a time-invariant attack.

Thus far, research on resilient control for CPS has provided most of the findings on DoS attacks. DoS attacks primarily cause the system to lose part of its information and render the state of the system unavailable. Different from the well-studied DoS attacks, the typical network anomalies, is also challenging for resilient control system design due to the easiness of anomalies implementation, and difficulties of anomalies detection, i.e., no system model knowledge is needed for network anomaly implementation, and network anomaly is stealthy based on healthy historical process data. In this case, it is hard to know the existence of a network anomaly in time, and thus resilient control against network anomaly is vital (for example, the well-known Stuxnet event \cite{a3,a4}). Moreover, for interconnected systems, spurious data due to network anomalies will propagate to subsystems through physical-level and cyber-level interconnections. Therefore, network anomalies significantly challenge the security of the entire CPS.

The literature, reports some representative control system design methods against network anomalies. For example, in \cite{a21} a Bernoulli distribution is considered to represent the probability of a successful replay attack, and the observer and controller of time-varying CPS are designed by solving recursive linear matrix inequalities to achieve consensus and $H_\infty$ performance of the closed-loop system. In \cite{a22}, by introducing dynamic variables and dwell time, an event-triggered communication scheme is proposed to reduce the network resource occupancy of multi-agent systems in the case of replay attacks, and its controller can also ensure the consensus and security of second-order systems under event triggering and attacks. In \cite{a23}, the formation tracking problem of multiple unmanned aerial vehicles under replay attacks is studied by using the sliding mode control method under switching event triggering scheme. In \cite{a24}, the replay attack is detected and identified, and the controller is analyzed and designed by identifying the attacker and detecting the sensors corresponding to the attack. In \cite{a25}, an elegant result on the resilient adaptive control of a kind of centralized nonlinear systems under replay attack is studied with system resilience quantification. However, the attack propagation and modeling uncertainty have not been addressed yet.

In \cite{a21}, the occurrence of network anomalies is described probabilistically, and the observer is designed for interconnected linear systems. In \cite{a22}, replay attacks are detected based on time stamp analysis, while \cite{a23} and \cite{a24} focus on mitigating adverse effects by addressing the detected replay attacks. The existences of a uniform upper bound on the active time and an accurate detection outcome of network anomalies are assumed in the literature above. However, these uniform upper bounds are not always available before the design of the controller and the accurate detection of network anomalies is challenging, where faking time stamp of data can be calculated by analyzing the eavesdropped data. Under this circumstance, the resilience of control systems is important due to the difficulties for knowing the presence of network anomalies. The analysis in \cite{a25} addresses the resilience conditions of centralized systems. However, when dealing with large-scale interconnected systems, these systems become more susceptible to network anomalies, and the complexity of the analysis increases significantly. This heightened vulnerability is a crucial concern that warrants careful consideration. Recalling the risk induced by network anomalies in the interconnected systems and considering that network anomalies are highly stealthy and difficult to detect with sufficient accuracy, it becomes interesting and significant to investigate how long the entire interconnected system can ``survive" without knowing how long, when, and which subsystem is under anomalies. Note that for the interconnected systems, the resilient control against network anomalies is still an open problem to be solved. The key challenges lie in the anomalies effect propagation, system uncertainty, and resilience quantification.

\subsection{Objectives and Contributions}

Motivated by the state-of-the-art, this work focuses on the resilient control problem of uncertain nonlinear interconnected systems under network anomalies. Differently from the reported results \cite{a21,a22,a23,a24}, neither accurate anomaly detection normal duration bound/model is assumed. Compared with \cite{a25}, we consider the interconnected system with complex anomaly effect propagation and uncertainty, and investigate the adaptive learning control issue. Considering the stealthiness's of network anomalies, the control system resilience is quantified by designing explicit boundary condition under the action of network anomalies. We adopt the distributed control strategy for the interconnected systems, we design the controller based on a back-stepping approach, and we approximate the model uncertainties by using neural networks.

The main contributions of the paper are given below.

\begin{itemize}
  \item In the framework of neural network approximation, the distributed controller and parameter estimator are designed by using the back-stepping method to enable the system resilience against network anomalies, and the stability of the anomalous and normal subsystems are analyzed.
  \item The quantitative resilience conditions for uncertain nonlinear interconnected systems under both local and neighboring network anomalies are provided in terms of the duration and resting time of each anomaly, ensuring that each subsystem remains asymptotically stable or bounded.
\end{itemize}

This paper is organized as follows: In Section~\ref{s2}, the problem to be solved is formulated. In Section~\ref{s3}, we give the controller and estimator design. In Section~\ref{s4}, the control system resilience is quantified. In Section~\ref{s5}, simulation results are provided to demonstrate the effectiveness of the theoretical analysis. Finally, some concluding remarks are given in Section~\ref{s6}.

\textbf{Notations}: For vectors $x_1$ and $x_2$, $col\left\{ x_1,x_2 \right\} =\left[ x_{1}^{T}\ x_{2}^{T} \right] ^T$. For vector $\theta$, $\lVert \theta \rVert$ denotes its Euclidean norm. $\mathcal{R}$ is the set of real numbers. $Z^+$ is the set of positive integers. $I$ is the unity matrix with appropriate dimension. For simplicity, we omit the function time variable $t$ or  state variable $x$ if the context is clear.

\section{Assumptions and Problem Formulation}\label{s2}

\subsection{Model description}
We consider a class of uncertain nonlinear dynamic systems, which include $ M $ interconnected subsystems. The $ i $-th subsystem is described as
\begin{equation}\label{1}
  \left\{ \begin{array}{l} 	\dot{x}_{i,j}\hspace{0.4em}=x_{i,j+1},\,\,\,\,\,\,j=1,...,n_i-1,\\ 	\dot{x}_{i,n_i}=\alpha _i\left( x_i \right) +u_i+\eta _i\left( x_i,t \right) +\beta _i\left( \bar{x}_i \right) ,\\ \end{array} \right.
\end{equation}
where $ i\in \mathcal{M}=\left\{ 1,...,M \right\}  $, $ x_i=col\left\{ x_{i,1},...,x_{i,n_i} \right\} \in \mathcal{R}^{n_i} $ is the state vector of the $i$-th subsystem, and $ \bar{x}_i\in \mathcal{R}^{\bar{n}_i} $ is the interconnected state vector of the $i$-th subsystem, including all the states of the adjacent subsystems that can affect the $i$-th subsystem. $ u_i\in \mathcal{R} $ is the control input of the $i$-th subsystem.
 $ \alpha _i\left( x_i \right) :\mathcal{R}^{n_i}\rightarrow \mathcal{R} $ represents the nominal model dynamics, and $ \eta _i\left( x_i,t \right) :\mathcal{R}^{n_i}\times \mathcal{R}^+\rightarrow \mathcal{R} $ represents the unknown modeling uncertainty associated with
 the $i$-th subsystem. Define the set $ \mathcal{N}_i\subset \mathcal{M} $, which represents the neighboring subsystems that can affect the $i$-th subsystem.
 $ \beta _i\left( \bar{x}_i \right) :\mathcal{R}^{\bar{n}_i}\rightarrow \mathcal{R} $ is the known interconnection function that satisfies
\begin{equation}\label{2}
  \beta _i\left( \bar{x}_i \right) =\sum_{j\in \mathcal{N}_i}^{}{\chi _{i,j}\left( x_j \right)}
\end{equation}
where $ \chi _{i,j}:\mathcal{R}^{n_j}\rightarrow \mathcal{R} $ are known functions.

\newtheorem{assumption}{Assumption}
\begin{assumption}
\rm{$ \alpha _i\left( x_i \right) $ and $\beta _i\left( \bar{x}_i \right)$ are locally Lipschitz continuous functions that satisfy the following property\cite{a26}:
\begin{align*}
&\left| \alpha_i\left( x_i \right) -\alpha_i\left( y_i \right) \right| \\
&\le \mathbb{L}_{\alpha_i}\left( \left| x_{i,1}-y_{i,1} \right| +...+\left| x_{i,n_i}-y_{i,n_i} \right| \right)
\end{align*}
\begin{align*}
&\left| \beta_i\left( \bar{x}_i \right) -\beta_i\left( \bar{y}_i \right) \right|\\
&\le \mathbb{L}_{\beta_i}\left( \left| x_{j,1}-y_{j,1} \right| +...+\left| x_{j,n_j}-y_{j,n_j} \right| \right)
\end{align*}
where $ \mathbb{L}_{\alpha_i} $ and $ \mathbb{L}_{\beta_i} $ are the Lipschitz constants for $\alpha_i\left( x_i \right)$ and $\beta_i\left( \bar{x}_i \right)$, respectively.}
\end{assumption}

\subsection{Network anomaly}
In order to quantify the resilience of the interconnected system and the bounded energy of the anomalies, we introduce the anomaly occurrence time series $t_h\left(h\in Z^+\right)$. Without loss of generality, $ t_0=0 $ is the system initial time, and we assume that the $h$-th anomaly start at the time $t_h$. $ t_{d,h} $ is the duration of the $h$-th anomaly. Since each anomaly launched by the anomaly initiator has finite energy, $ t_{d,h} $ is bounded \cite{a22}. Hence, the $h$-th network anomaly occurs at $ t\in \left[ t_h,t_h+t_{d,h} \right)  $. The resting time after the $ h $-th anomaly is defined as $ t_{r,h}=t_{h+1}-\left( t_h+t_{d,h} \right)  $, which indicates the duration between two consecutive anomalies.
\newtheorem{remark}{\textbf{Remark}}
\begin{remark}
\rm{As the anomalous subsystems and scenarios remain unknown, the overall resilience of the interconnected systems is the major concern, which means that -- for resilience -- a boundary condition on a worst case scenario has to be designed. Thus, without loss of generality, the anomaly occurrence time series and duration are not distinguished with respect to subsystems to simplify the analysis. If the anomaly scenarios can be identified, the time series and duration of anomaly related to specific subsystems can be noted based on the current result easily.}
\end{remark}

\begin{figure}[htbp]
 \centering
 \includegraphics[width=3.3in]{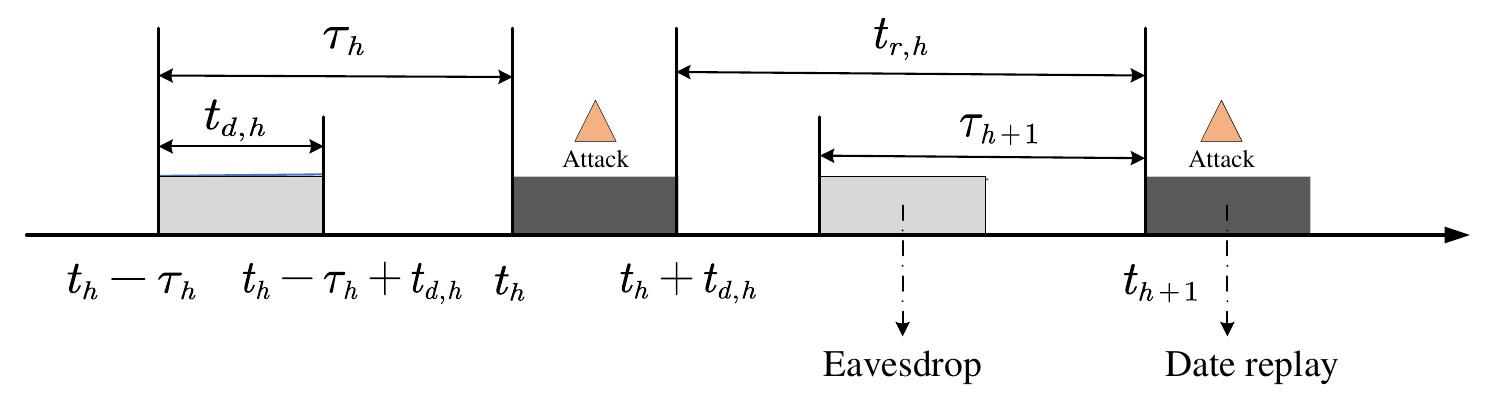}
 \caption{Time series of network anomaly}
 \vspace{-1mm}
 \label{Fig1}
\end{figure}

The network anomaly occurs in two stages, the anomaly initiator eavesdrops system data at $ t\in \left[ t_h-\tau _h,t_h+t_{d,h}-\tau _h \right)  $ in advance. $ \tau _h $ is a randomly selected constant for the $h$-th anomaly. \cite{a25}.
Then, when the anomaly initiator launches a network anomaly, the pre-obtained state information is transmitted in place of the current state information, such that the system can not acquire the real state $x\left( t \right)$, but the previously retained historical state $ x\left( t-\tau _h \right)$, as shown in~Fig. \ref{Fig1}.
\subsection{Adaptive approximation}

A distributed control scheme is considered for the interconnected system, and we use the information of local and adjacent subsystems to design the corresponding distributed controller.

To deal with the unknown modeling uncertainty, the adaptive approximation scheme is adopted, and a linearly parameterized approximator $ \hat{\eta}_i\left( x_i,t \right)=\theta _{i}^{T}\pi _i\left( x_i \right)  $ is introduced as the approximation of $\eta_i\left( x_i,t \right)$, where $ \theta _{i}^{}\in\mathcal{R}^{p_i}$ is an unknown vector of network weights and $ \pi_i\left( x_i \right) :\mathcal{R}^{n_i}\rightarrow \mathcal{R}^{p_i} $ represent the known basis function vector.
\begin{assumption}
\rm{$ \pi _i\left( x_i \right) $ is a Lipschitz continuous function vector that satisfy the following property:
\begin{align*}
\lVert \pi_i\left( x_i \right) -\pi_i\left( y_i \right) \rVert\le \mathbb{L}_{\pi_i} \lVert x_{i}-y_{i} \rVert
\end{align*}
where $ \mathbb{L}_{\pi_i} $ is a known Lipschitz constant.}
\end{assumption}

As $\pi _i\left( x_i \right) $ is chosen by the designer for uncertainty characterization, Assumption 2 is easy to meet (e.g., radial basis function is adopted as the basis function). To quantify the approximation performance, define the optimal weight vector
\begin{align*}
\theta _{i}^{*}=\text{arg}\left\{ \underset{\theta _i\in \mathcal{R}^{p_i}}{\text{inf}}\left\{ \underset{x_i\in \mathcal{X}_{i}}{\text{sup}}\left| \eta _i\left( x_i,t \right) -\hat{\eta}_i\left( x_i,t \right) \right| \right\} \right\}
\end{align*}
where $\mathcal{X}_{i}\in\mathcal{R}^{n_i}$ is the concerned state space.
Based on  $ \theta _{i}^{*}$, the gap between $\eta _i\left( x_i,t \right)$ and $\hat{\eta}_i\left( x_i,t \right)$ corresponds to the following assumption.

\begin{assumption}
\rm{For $i\in{ \mathcal{M}}$ and $x_i\in \mathcal{X}_{i}$, assume that $ \left| \eta _i\left( x_i,t \right) -\hat{\eta}_i\left( x_i,t \right) \right|\le \lambda _{i}^{*}\varOmega _i\left( x_i \right)  $, where $ \lambda _{i}^{*}\ge 0 $ is an unknown constant and $ \varOmega _i\left( x_i \right):\mathcal{R}^{n_i}\rightarrow \mathcal{R}^{+}  $ is the known bounding function.}
\end{assumption}

To simplify the analysis, assume that $ \lVert \theta _{i}^{*} \rVert <\bar{\theta} $ and $ \left| \lambda _{i}^{*} \right|<\bar{\lambda} $ for $i\in{ \mathcal{M}}$. To quantify the resilience conditions, $\bar{\theta}$ and $\bar{\lambda} $ need to be known.
\begin{assumption}
\rm{$ \varOmega_i\left( x_i \right) $ is a Lipschitz continuous function that satisfy the following property:
\begin{align*}
&\left| \varOmega_i\left( x_i \right) -\varOmega_i\left( y_i \right) \right| \\
&\le \mathbb{L}_{\varOmega_i}\left( \left| x_{i,1}-y_{i,1} \right| +...+\left| x_{i,n_i}-y_{i,n_i} \right| \right)
\end{align*}
where $ \mathbb{L}_{\varOmega_i} $ is a known Lipschitz constant.}
\end{assumption}

$\varOmega _i\left( x_i \right) $ is used to model an upper bound on the approximation error, and the assumed Lipschitz condition is easily satisfied.

\begin{remark}
 In this study, the local Lipschitz condition is assumed for system dynamics. First, as we employed the adaptive approximation scheme in the study, it generally describes an approximation error bound in terms of a concerned state space \cite{a27}, e.g., $ x_{i} \in \mathcal{X}_{i}$ and $\mathcal{X}_{i}\in\mathcal{R}^{n_i}$. Second, due to the existence of network anomalies, one cannot know the exact space the real-time state belongs to, even if we introduce multiple local Lipschitz constants or piecewise description for system dynamics. To quantify the control performance and match with the adaptive approximation scheme, we employ this ``lumped" Lipschitz condition in Assumption 1. All in all, the locality of Lipschitz condition as well as the state space $ \mathcal{X}_{i}$ only relates to the state space we concerned, not a pre-assumption about the boundedness of state. Moreover, the convergence and the obtained resilience boundary condition of the control system are derived without assuming the boundedness of system state.
\end{remark}
\subsection{Problem statement}
\begin{figure}[htbp]
 \centering
 \includegraphics[width=3.3in]{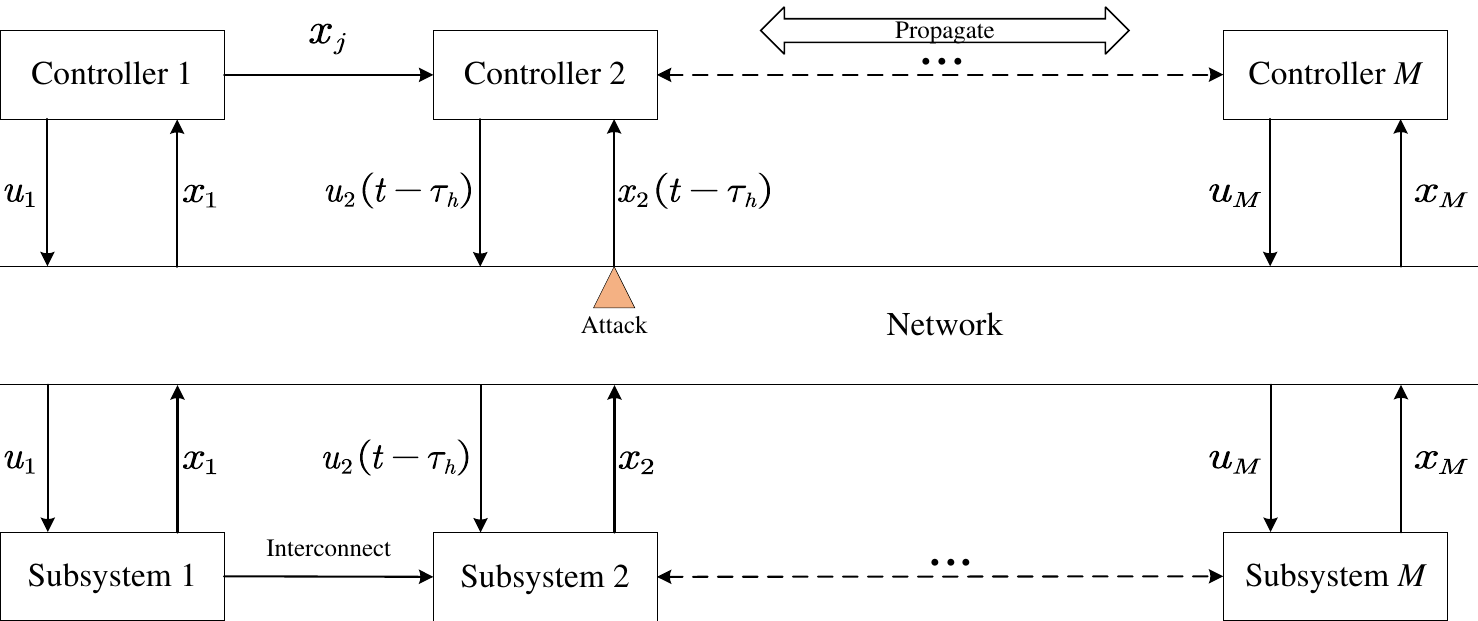}
 \caption{The control structure of the interconnected system}
 \vspace{-1mm}
 \label{Fig2}
\end{figure}

In the interconnected system, there are physical interconnections between subsystems. When a subsystem experiences a network anomaly, the state of the subsystem changes, which affects the local control loop. Note that the information interaction between the controllers also leads to the propagation of the anomaly, as shown in~Fig. \ref{Fig2}.

Specifically, as illustrated in~Fig. \ref{Fig2}, the anomaly occurs during the network transmission of signals. When a subsystem is subjected to a network anomaly, the controller receives historical delayed state information for real-time control signal calculation. Furthermore, this historical state information propagates to adjacent subsystems through cyber-physical information interactions.

Based on the aforementioned control system description, our two main objectives are:
\begin{enumerate}
  \item How to design the controller and parameter estimator under network anomalies, and ensure stability of the entire interconnected system.
  \item When multiple anomalies are considered and the topology of the anomalous subsystems are unknown, the resilience conditions of the anomalous control systems are quantified with respect to the anomaly time and the resting time.
\end{enumerate}

\section{Dessign of the distributed adaptive controller}\label{s3}

First of all, we present the distributed controller and parameter estimator design when all subsystems are normal, combining back-stepping and adaptive approximation techniques\cite{a27}.

For distributed controller design, a new state vector is defined recursively by the following coordinate transformation:
\begin{equation}\label{3}
  \left\{ \begin{array}{l} 	z_{i,j}\hspace{0.4em}=x_{i,j}-\alpha _{i,j-1},\,\,\,\,j=1,...,n_i-1,\\ 	
  z_{i,n_i}=x_{i,n_i}-\alpha _{i, n_i-1} ,\\ \end{array} \right.
\end{equation}
where $ z_{i,j} $ is the virtual control error and $ \alpha _{i, j-1 } $ is the virtual control of $ x_{i,j} $. The design of the distributed adaptive controller is recursive in the sense that the computation of $ \alpha _{i,j} $ relies on computing $ \alpha _{i,j-1  } $\cite{a28}.

\text{Step $1$}:
Let $ \alpha _{i,0}=0 $. Using~(\ref{1}) and the change of coordinate transformation given by~(\ref{3}), one obtains
\begin{equation}\label{4}
\dot{z}_{i,1}=z_{i,2}+\alpha _{i,1}
\end{equation}
Let the virtual controller $ \alpha _{i,1}=-\gamma z_{i,1} $ and choose the first Lyapunov function $ V_{i,1}=\frac{1}{2}z_{i,1}^{2} $, where $ \gamma >1 $ is the control gain determined by the designer, and then it comes
\begin{equation}\label{5}
\dot{z}_{i,1}=-\gamma z_{i,1}+z_{i,2}
\end{equation}
Based on~(\ref{5}), the time-derivative of $ V_{i,1} $ is given by
\begin{equation}\label{6}
\dot{V}_{i,1}=-\gamma z_{i,1}^{2}+z_{i,1}z_{i,2}
\end{equation}

\text{Step $ j\left( 2\le j\le n_i-1 \right)  $}:
A similar procedure is employed recursively at each step $ j=2,3,...,n_i-1 $.
The time-derivative of $ z_{i,j} $ is
\begin{equation}\label{7}
\dot{z}_{i,j}=z_{i, j+1  }+\alpha _{i,j}-\sum_{k=1}^{j-1}{\frac{\partial \alpha _{i, j-1  }}{\partial x_{i,k}}x_{i, k+1  }}
\end{equation}
The virtual controller is designed as
\begin{equation}\label{8}
\alpha _{i,j}=-\gamma z_{i,j}-z_{i, j-1  }+\sum_{k=1}^{j-1}{\frac{\partial \alpha _{i, j-1  }}{\partial x_{i,k}}x_{i, k+1  }}
\end{equation}
and the Lyapunov function is
\begin{equation}\label{9}
V_{i,j}=V_{i,j-1}+\frac{1}{2}z_{i,j}^{2}
\end{equation}
Taking~(\ref{8}) into~(\ref{7}), one obtains
\begin{equation}\label{10}
\dot{z}_{i,j}=-\gamma z_{i,j}-z_{i, j-1  }+z_{i, j+1  }
\end{equation}
Based on~(\ref{9}) and~(\ref{10}),  the time-derivative of $ V_{i,j} $ is given by
\begin{equation}\label{11}
\dot{V}_{i,j}=-\sum_{k=1}^j{\gamma z_{i,k}^{2}}+z_{i,j}z_{i, j+1  }
\end{equation}

\text{Step $ n $}:
In the final design step, the actual control input $ u_i $ appears.
The time-derivative of $ z_{i,n_i} $ is
\begin{align}\label{12}
  \dot{z}_{i,n_i}=\ &\alpha _i\left( x_i \right) +u_i+\eta _i\left( x_i,t \right) +\beta _i\left( \bar{x}_i \right) \nonumber\\
   &-\sum_{k=1}^{n_i-1}{\frac{\partial \alpha _{i, n_i-1  }}{\partial x_{i,k}}x_{i, k+1 }}
\end{align}
The actual control is designed as
\begin{align}\label{13}
 \left\{ \begin{array}{l} 	\bar{u}_i\hspace{0.6em}=\bar{u}_{i,1}+\bar{u}_{i,2}+\bar{u}_{i,3}\\ 	\bar{u}_{i,1}=-\gamma z_{i,n_i}-z_{i,n_i-1}-\alpha _i\left( x_i \right) -\beta _i\left( \bar{x}_i \right) \\\hspace{2.9em}+\sum_{k=1}^{n_i-1}{\frac{\partial \alpha _{i,n_i-1}}{\partial x_{i,k}}x_{i,k+1}}\\ 	 	\bar{u}_{i,2}=-\hat{\theta}_{i}^{T}\pi _i\left( x_i \right)\\ 	\bar{u}_{i,3}=-\hat{\lambda}_i\varOmega _i\left( x_i \right)\\ \end{array} \right.
\end{align}
where $ \hat{\theta}_i $ and $ \hat{\lambda}_i $ are the estimations of $ \theta _{i}^{*}$  and $ \lambda _{i}^{*}$, respectively. Also $ \lVert \hat\theta _{i}^{} \rVert <\bar{\theta} $ and $ \left|\hat \lambda _{i}^{} \right|<\bar{\lambda} $ are hold. For example, it is often possible to use projection to keep the estimated parameters within a certain range.

We consider the Lyapunov function
\begin{equation}\label{14}
V_{i,n_i}=V_{i,n_i-1 }+\frac{1}{2}z_{i,n_i}^{2}+\frac{1}{2}\tilde{\theta}_{i}^{T}\varGamma _{i}^{-1}\tilde{\theta}_i+\frac{1}{2\zeta _i}\tilde{\lambda}_{i}^{2}
\end{equation}
where $\zeta _i > 0$ and $\varGamma _{i}$ is positive definite, $ \tilde{\theta}_i=\theta _{i}^{*}-\hat{\theta}_i $ represents the network weight estimation error and $ \tilde{\lambda}_i=\lambda _{i}^{*}-\hat{\lambda}_i $ represents the bounding parameter error.

Taking~(\ref{13}) into~(\ref{12}), one obtains
\begin{equation}\label{15}
 \dot{z}_{i,n_i}\le-\gamma z_{i,n_i}-z_{i,n_i-1  }+\tilde{\theta}_{i}^{T}\pi _i\left( x_i \right) +\tilde{\lambda}_i\varOmega _i\left( x_i \right)
\end{equation}
According to Assumption 3 and~(\ref{14}), the time-derivative of $ V_{i,n_i} $ is given by
\begin{align}\label{16}
\dot{V}_{i,n_i}\le &-\sum_{k=1}^{n_i}{\gamma z_{i,k}^{2}}+z_{i,n_i}\left( u_i-\bar{u}_{i,1} \right)\nonumber\\
&+z_{i,n_i}\left(\pi^{T} \left( x_i \right)\theta _{i}^{*} +\lambda _{i}^{*}\varOmega \left( x_i \right) \right)\nonumber\\
&-\tilde{\theta}_{i}^{T}\varGamma _{i}^{-1}\dot{\hat{\theta}}_i-\tilde{\lambda}_i\zeta _{i}^{-1}\dot{\hat{\lambda}}_i
\end{align}
The controller and parameter estimator are designed as
\begin{equation}\label{17}
\left\{ \begin{array}{l} 	u_i=\bar{u}_i\left( t \right)\\ 	
\dot{\hat{\theta}}_i\hspace{0.1em}= \varphi _i\left( t \right) \\
\dot{\hat{\lambda}}_i= \phi _i\left( t \right)\\
 \end{array} \right.
\end{equation}
where $ \varphi _i=\varGamma z_{i,n_i}\pi _i\left( x_i \right)  $ and $ \phi _i=\zeta _iz_{i,n_i}\varOmega _i\left( x_i \right)  $. Since we only consider the concerned state space, for simplicity, the projection operator omitted here.
Then, $\dot{V}_{i,n_i}\le -\sum_{k=1}^{n_i}{\gamma z_{i,k}^{2}}$, which implies that the uncertain nonlinear interconnected system controlled by~(\ref{17}) is stable.

In summary, one can conclude that
\begin{equation}\label{18}
\left\{ \begin{array}{l} 	\dot{z}_{i,1}\hspace{0.4em}=-\gamma z_{i,1}+z_{i,2}\\ 	
\dot{z}_{i,j}\hspace{0.4em}=-\gamma z_{i,j}-z_{i, j-1  }+z_{i, j+1  }\\
\dot{z}_{i,n_i}\le-\gamma z_{i,n_i}-z_{i,n_i-1  }+\tilde{\theta}_{i}^{T}\pi _i\left( x_i \right) \\ \hspace{3.25em}+\tilde{\lambda}_i\varOmega _i\left( x_i \right)\\
 \end{array} \right.
\end{equation}

By combining~(\ref{3}),~(\ref{8}), and~(\ref{13}), each $ x_{i,j} $ can be represented by a linear combination of $ z_{i,j} $ and satisfies \cite{s3},
\begin{equation}\label{19}
  \left\{ \begin{array}{l} 	x_{i,j}=c_{i,j,1}z_{i,1}+\cdots +c_{i,j,j}z_{i,j}\\
  	\bar{u}_i\hspace{0.65em}=-\gamma z_{i,n_i}-z_{i, n_i-1 }-\alpha _i\left( x_i \right)-\beta _i\left( \bar{x}_i \right)\\
 	\hspace{2.9em}+b_{i,n_i,1}z_{i,1}+\cdots +b_{i,n_i,n_i}z_{i,n_i} \\
 \hspace{2.9em}-\hat{\theta}_{i}^{T}\pi _i\left( x_i \right) -\hat{\lambda}_i\varOmega _i\left( x_i \right)\\ \end{array} \right.
\end{equation}
where $c_{i,j,j}$ and $b_{i,n_i,j} $ are constants that can be computed based on the dynamics of each subsystem.

\section{Resilience quantification with network anomalies}\label{s4}
In this section, the qualitative and quantitative  resilience conditions of the interconnected control system under network anomalies are derived. Due to the presence of network anomaly, the anomalous subsystems cannot access the real-time states, where the controller and parameter estimator of the anomalous subsystems receive only the delayed feedback information. These tampered states propagate to other neighboring subsystems via distributed control (cyber interconnection). As multiple anomalies, multiple subsystems, and interconnections are considered, the health status of each subsystem differs. To better describe the health status of subsystems, we give the following definition of sets.

\begin{figure}[htbp]
 \centering
 \includegraphics[width=3.3in]{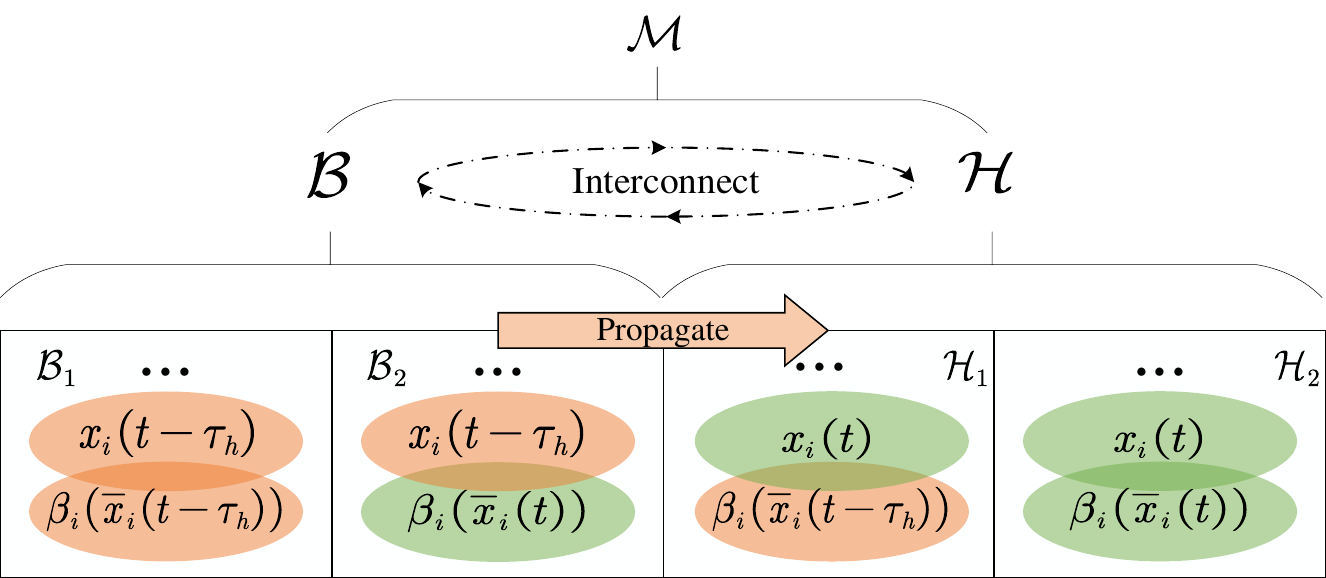}
 \caption{The received states for the distributed controllers of the nonlinear interconnected system under multiple network anomalies. (The pale orange color represents the historical state that has been altered due to the network anomaly, while the pale green color indicates a secure state.)}
 \vspace{-1mm}
 \label{Fig3}
\end{figure}

We classify all subsystems into four distinct sets, considering the perspectives of network anomalies on both the subsystems themselves and the interconnected functional components. We define set $ \mathcal{B}\subseteq \mathcal{M} $ for denoting all the anomalous subsystems and set $\mathcal{H}=\left\{ i\left| i\in \mathcal{M}\ \setminus\ \right.\mathcal{B} \right\}$ that represents all normal subsystems. Due to the interconnection, the anomaly effect can propagate with the information for distributed control. Thus, the sets $\mathcal{B}$ and $\mathcal{H}$ are all split into two parts. As shown in~Fig. \ref{Fig3}, $ \mathcal{B}_1 $ and $ \mathcal{H}_1 $ are sets for denoting these anomalous and normal subsystems with anomalous neighboring subsystems, respectively;  $ \mathcal{B}_2 $ and $ \mathcal{H}_2 $ are sets for denoting these anomalous and normal subsystems without any anomalous neighboring subsystems, respectively.

It is particularly important to note that $\beta _i\left( \bar{x}_i\left( t-\tau _h \right) \right)$ represents the interconnection function influenced by the historical states of neighboring subsystems. However, we have used a simplified notation here. In reality, the interconnection function $\beta _i$ may include different neighboring states, some of which may have been affected by network anomaly, while others may not have been. The specific circumstances will be discussed in the following analysis.

For the resilience quantification, we split the analysis into two scenarios, the time duration of each anomaly and resting time between two consecutive anomalies.
\subsection{The time duration of each anomaly}
Since the security of the entire control system is considered, the lower bound of the system resilience against network anomaly is expected. Thus, we will consider how long the duration of the anomalies the system can tolerate. Regarding the time duration of each anomaly, the major result is displayed in the following theorem.
\newtheorem{theorem}{\textbf{Theorem}}
\begin{theorem}
Under Assumptions 1-4 and $\gamma>1$, the uncertain nonlinear interconnected system given by~(\ref{1}) and controlled by~(\ref{17}) is asymptotically stable, if the duration of each network anomaly satisfies
\begin{equation}\label{20}
t_{d,h}\le \frac{1}{w_1-w_2}ln\frac{w_1w_3+w_1}{w_2w_3+w_1}
\end{equation}

\begin{equation}\label{21}
 \left\{ \begin{array}{l} 	w_1=c_m\left( \mathbb{L}_{\alpha}+3\bar\theta\mathbb{L}_{\pi}+3\bar\lambda\mathbb{L}_{\varOmega}+M\mathbb{L}_{\beta} \right) \\
 \hspace{2.5em}+3\gamma +4+b_m\\ 	
 w_2=c_m\left( \mathbb{L}_{\alpha}+3\bar\theta\mathbb{L}_{\pi}+3\bar\lambda\mathbb{L}_{\varOmega}+M\mathbb{L}_{\beta} \right) \\
 \hspace{2.5em}+\gamma +b_m\\ 	
w_3=\frac{\sqrt{w_{2}^{2}+\frac{8c_m\left( \bar{\lambda}\mathbb{L}_{\varOmega}+\bar{\theta}\mathbb{L}_{\pi} \right)}{Mn_m}}-w_2}{4c_m\left( \bar{\lambda}\mathbb{L}_{\varOmega}+\bar{\theta}\mathbb{L}_{\pi} \right)}\\
 \end{array} \right.
\end{equation}
where $n _m=\max \left\{n_i :i=\left\{ 1,...,M \right\} \right\}$, and
\begin{equation}\label{22}
	 \mathbb{L}_{*}\hspace{0.2em}=\max \left\{ \mathbb{L}_{*_i}:i=1,...,M \right\}\hspace{1em}
({*}=\alpha,\beta,\pi,\varOmega)
\end{equation}
\begin{equation}\label{23}
 \left\{ \begin{array}{l} 	b_{i,m}=\max \left\{ \left| b_{i,n_i,1} \right|,...,\left| b_{i,n_i,n_i} \right| \right\}\\
	b_m\hspace{0.5em}=\max \left\{ \left| b_{i,m} \right|:i=\left\{ 1,...,M \right\} \right\}\\ \end{array} \right.
\end{equation}
\begin{equation}\label{24}
 \left\{ \begin{array}{l} 	\bar{c}_{i,1}\hspace{0.3em}=\sum_{j=1}^{n_i}{\left| c_{i,j,1} \right|},...,\bar{c}_{i,n_i}=\sum_{j=n_i}^{n_i}{\left| c_{i,j,n_i} \right|}\\
 	c_{i,m}=\max \left\{ \bar{c}_{i,1},...,\bar{c}_{i,n_i} \right\}\\
 	c_m\hspace{0.5em}=\max \left\{ c_{i,m}:i=\left\{ 1,...,M \right\} \right\}\\ \end{array} \right.
\end{equation}

\end{theorem}
~\\
\textbf{Proof}: First, anomalous subsystem and controller changes are quantified and discussed for $ i\in \mathcal{B}$ and $i\in\mathcal{H}$, respectively. Finally, the quantitative resilient condition is analyzed for the entire interconnected system.

\textbf{Part\ 1:} Anomalous subsystem

Without loss of generality, let's consider $ i\in \mathcal{B} $, which means that the $ i $-th subsystem is under anomaly. As the newest states of the $i$-th subsystem cannot be obtained due to the $h$-th anomaly, the control signal and parameter estimator turn to be
\begin{equation}\label{25}
\left\{ \begin{array}{l} 	u_i=\bar{u}_i\left( t-\tau _h \right)\\ 	
\dot{\hat{\theta}}_i\hspace{0.1em}=\varphi _i\left( t-\tau _h  \right)\\
\dot{\hat{\lambda}}_i= \phi _i\left( t-\tau _h  \right)\\
 \end{array} \right.
\end{equation}
Taking~(\ref{25}) into~(\ref{16}), the Lyapunov function becomes
\begin{align}\label{26}
\dot{V}_{i,n_i}\le& -\sum_{k=1}^{n_i}{\gamma z_{i,k}^{2}}+z_{i,n_i}\left( \bar{u}_i\left( t-\tau _h \right) -\bar{u}_{i,1} \right) \nonumber \\
&+z_{i,n_i}\left( \pi^{T} \left( x_i \right)\theta _{i}^{*} +\lambda _{i}^{*}\varOmega \left( x_i \right) \right)\nonumber \\
&-\tilde{\theta}_{i}^{T}\varGamma _{i}^{-1}\varphi _i\left( t-\tau _h \right) -\tilde{\lambda}_i\zeta _{i}^{-1}\phi _i\left( t-\tau _h \right)
\end{align}

Furthermore, for the neighboring subsystems of subsystem $i$, we define $ \mathcal{G}_i=\mathcal{N}_i\cap \mathcal{B}$ and $\mathcal{F}_i=\mathcal{N}_i\cap \mathcal{H} $ to indicate the anomalous and normal neighboring subsystem sets, respectively. For $i\in\mathcal{B}$, the interconnection function for the $i$-th controller turns to be
\begin{align}\label{27}
\beta _i\left( \bar{x}_i\left( t-\tau _h \right) \right) =&\sum_{j\in \mathcal{G}_i}^{}{\chi _{i,j}\left( x_j\left( t-\tau _h \right) \right)}\nonumber \\
&+\sum_{j\in \mathcal{F}_i}^{}{\chi _{i,j}\left(x_j\left(t\right)\right)}
\end{align}
Specifically, if $i\in\mathcal{B}_1$, $\mathcal{G}_i\neq \varnothing$; if $i\in\mathcal{B}_2$, $\mathcal{G}_i=\varnothing$. Based on~(\ref{26}), the interconnection discrepancy due to network anomaly for the distributed controller is given by
\begin{align}\label{28}
 &\hspace{0.5em}\beta _i\left( \bar{x}_i \right) -\beta _i\left( \bar{x}_i\left( t-\tau _h \right) \right) \nonumber\\
 &=\sum_{j\in \mathcal{G}_i}^{}{\left( \chi _{i,j}-\chi _{i,j}\left( x_j\left( t-\tau _h \right) \right) \right)}
\end{align}

Divide~(\ref{26}) into the following three formulas to analyze the discrepancy.

First, the anomalous control effort discrepancy for $\bar{u}_{i,1}$ is obtained as
\begin{equation}\label{29}
z_{i,n_i}\left( \bar{u}_{i,1}\left( t-\tau _h \right) -\bar{u}_{i,1} \right)
\end{equation}

Second, the estimator $\hat\theta_i$ cause the Lyapunov function discrepancy is obtained as
\begin{align}\label{30}
&z_{i,n_i}\left( \bar{u}_{i,2}\left( t-\tau _h \right)+\pi^{T}\left( x_i \right)\theta _{i}^{*} \right)\nonumber\\
&-\tilde{\theta}_{i}^{T}\varGamma _{i}^{-1}\varphi _i\left( t-\tau _h \right)
\end{align}

Third, the estimator $\hat\lambda_i$ cause the Lyapunov function discrepancy is is obtained as
\begin{align}\label{31}
&z_{i,n_i}\left( \bar{u}_{i,3}\left( t-\tau _h \right) +\lambda _{i}^{*}\varOmega \left( x_i \right) \right)\nonumber\\
&-\tilde{\lambda}_i\zeta _{i}^{-1}\phi _i\left( t-\tau _h \right)
\end{align}
Define
\begin{equation}\label{32}
\left\{ \begin{array}{l} z_i\hspace{0.7em}=\left| z_{i,1} \right|+...+\left| z_{i,n_i} \right|	\\ 	
e_{i,h}=\sum_{k=1}^{n_i}{\left| z_{i,k} -z_{i,k}\left( t-\tau _h \right) \right|}\\
e_h\hspace{0.6em}=\sum_{i=1}^M{\sum_{k=1}^{n_i}{\left| z_{i,k}-z_{i,k}\left( t-\tau _h \right) \right|}}\\ \end{array} \right.
\end{equation}
For ~(\ref{29}), from~(\ref{13}),~(\ref{23}) and~(\ref{32}), the anomalous control effort discrepancy for $\bar{u}_{i,1}$ is obtained as
\begin{align}\label{33}
\bar{u}_{i,1}\left( t-\tau _h \right) -\bar{u}_{i,1}\le &\mspace{5.4mu}\alpha _i\left( x_i \right) -\alpha _i\left( x_i\left( t-\tau _h \right) \right) \nonumber\\ 	
&+\beta _i\left( \bar{x}_i \right) -\beta _i\left( \bar{x}_i\left( t-\tau _h \right) \right) \nonumber\\
&+e_{i,h}\left( b_{i,m}+\gamma \right)
\end{align}
Then, analyzing the anomalous $\alpha _i $ and $\beta _i $ functions state discrepancy. According to the Assumption 1,~(\ref{19}),~(\ref{24}), and~(\ref{32}), one knows
\begin{align}\label{34}
\left| \alpha _i -\alpha _i\left( x_i\left( t-\tau _h \right) \right) \right|
&\le \mathbb{L}_{\alpha_i}\sum_{k=1}^{n_i}{\left| x_{i,k} -x_{i,k}\left( t-\tau _h \right) \right|}\nonumber\\
 &\le \mathbb{L}_{\alpha_i}\sum_{k=1}^{n_i}{\bar{c}_{i,k}\left| z_{i,k} -z_{i,k}\left( t-\tau _h \right) \right|}\nonumber\\
 &\le \mathbb{L}_{\alpha_i}c_{i,m}e_{i,h}
\end{align}
Also, according to Assumption 1,~(\ref{19}),~(\ref{24}), and~(\ref{32}), one knows
\begin{align}\label{35}
\left| \beta _i -\beta _i\left( \bar{x}_i\left( t-\tau _h \right) \right) \right|
&\le \mathbb{L}_{\beta_i}\sum_{j\in \mathcal{G}_i}^{}\sum_{k=1}^{n_j}{\left| x_{j,k} -x_{j,k}\left( t-\tau _h \right) \right|}\nonumber\\
&\le {\mathbb{L}_{\beta_i}\sum_{j\in \mathcal{G}_i}^{}c_{j,m}e_{j,h}}
\end{align}
Substitute~(\ref{34}) and~(\ref{35}) into~(\ref{33}), it yields
\begin{align}\label{36}
\bar{u}_{i,1}\left( t-\tau _h \right) -\bar{u}_{i,1}\le &\hspace{0.5em}e_{i,h}\left( b_{i,m}+\gamma+\mathbb{L}_{\alpha_i}c_{i,m} \right)\nonumber\\
&+\mathbb{L}_{\beta_i}\sum_{j\in \mathcal{G}_i}^{}{c_{j,m}e_{j,h}}
\end{align}
Therefore, for~(\ref{29}), the bound obtained as
\begin{align}\label{37}
z_{i,n_i}\left(\bar{u}_{i,1}\left( t-\tau _h \right) -\bar{u}_{i,1}\right)\le &\hspace{0.5em}z_ie_{i,h}\left( b_{i,m}+\gamma+\mathbb{L}_{\alpha_i}c_{i,m} \right)\nonumber\\
&+z_i\mathbb{L}_{\beta_i}\sum_{j\in \mathcal{G}_i}^{}{c_{j,m}e_{j,h}}
\end{align}
For~(\ref{30}), from~(\ref{13}) and~(\ref{17}), one knows
\begin{align}\label{38}
&z_{i,n_i}\left( \bar{u}_{i,2}\left( t-\tau _h \right) +\theta _{i}^{*T}\pi \left( x_i \right) \right)-\tilde{\theta}_{i}^{T}\varGamma _{i}^{-1}\varphi _i\left( t-\tau _h \right) \nonumber\\
&=z_{i,n_i}\left( \pi ^T\left( x_i \right) \theta _{i}^{*}-\hat{\theta}_{i}^{T}\left( t-\tau _h \right) \pi _i\left( x_i\left( t-\tau _h \right) \right) \right) \nonumber\\
&\hspace{1em}-\left( \theta _{i}^{*}-\hat{\theta}_i\left( t-\tau _h \right) \right) ^Tz_{i,n_i}\left( t-\tau _h \right) \pi _i\left( x_i\left( t-\tau _h \right) \right) \nonumber\\
&=\left( z_{i,n_i}\pi \left( x_i \right) -z_{i,n_i}\left( t-\tau _h \right) \pi _i\left( x_i\left( t-\tau _h \right) \right) \right) ^T\theta _{i}^{*}\nonumber\\
&\hspace{1em}+\left( z_{i,n_i}\left( t-\tau _h \right) -z_{i,n_i} \right) \hat{\theta}_{i}^{T}\left( t-\tau _h \right) \pi _i\left( x_i\left( t-\tau _h \right) \right)
\end{align}
Then, divide~(\ref{38}) into two terms,
\begin{equation}\label{39}
\left( z_{i,n_i}\pi \left( x_i \right) -z_{i,n_i}\left( t-\tau _h \right) \pi _i\left( x_i\left( t-\tau _h \right) \right) \right)^{T}\theta _{i}^{*}
\end{equation}
and
\begin{equation}\label{40}
\left( z_{i,n_i}\left( t-\tau _h \right) -z_{i,n_i} \right) \hat{\theta}_{i}^{T}\left( t-\tau _h \right) \pi _i\left( x_i\left( t-\tau _h \right) \right)
\end{equation}
Derive the bounds for~(\ref{39}) and~(\ref{40}), respectively.

In the first term~(\ref{39}), add and subtract the same term $z_{i,n_i}\pi _i\left( x_i\left( t-\tau _h \right) \right)^T
\theta _{i}^{*}$, keep unchanged formula as
\begin{align}\label{41}
&\left( z_{i,n_i}\pi \left( x_i \right) -z_{i,n_i}\left( t-\tau _h \right) \pi _i\left( x_i\left( t-\tau _h \right) \right) \right)^{T}\theta _{i}^{*}
\nonumber\\
&=\left(z_{i,n_i}\pi \left( x_i \right) -z_{i,n_i}\pi _i\left( x_i\left( t-\tau _h \right) \right) \right) ^T\theta _{i}^{*}
\nonumber\\
&\hspace{1em}+\left( z_{i,n_i}-z_{i,n_i}\left( t-\tau _h \right) \right) \pi _i\left( x_i\left( t-\tau _h \right) \right) ^T\theta _{i}^{*}
\end{align}
Analyzing the anomalous $\pi _i$ functions state discrepancy. According to Assumption 2,~(\ref{19}),~(\ref{24}) and~(\ref{32}), one knows
\begin{align}\label{42}
\lVert \pi _i -\pi _i\left( x_i\left( t-\tau _h \right) \right) \rVert &\le \mathbb{L}_{\pi _i} \lVert x_i-x_i\left( t-\tau _h \right) \rVert \nonumber\\
&\le \mathbb{L}_{\pi _i}\sum_{k=1}^{n_i}{\left| x_{i,k}-x_{i,k}\left( t-\tau _h \right) \right|}\nonumber\\
&\le \mathbb{L}_{\pi _i}\sum_{k=1}^{n_i}{\bar{c}_{i,k}\left| z_{i,k} -z_{i,k}\left( t-\tau _h \right) \right|}\nonumber\\
&\le \mathbb{L}_{\pi _i}c_{i,m}e_{i,h}
\end{align}
From~(\ref{32}) and~(\ref{42}), one obtains
\begin{equation}\label{43}
z_{i,n_i}\left(\pi \left( x_i \right) -\pi _i\left( x_i\left( t-\tau _h \right) \right) \right) ^T\theta _{i}^{*}\le z_i\bar{\theta}\mathbb{L}_{\pi _i}c_{i,m}e_{i,h}
\end{equation}
and
\begin{align}\label{44}
&\left( z_{i,n_i}-z_{i,n_i}\left( t-\tau _h \right) \right) \pi _i\left( x_i\left( t-\tau _h \right) \right) ^T\theta _{i}^{*}\nonumber\\
&\le \bar{\theta}\mathbb{L}_{\pi _i}e_{i,h}\sum_{k=1}^{n_i}{\left( \left| x_{i,k}\left( t \right) -x_{i,k}\left( t-\tau _h \right) \right|+\left| x_{i,k}\left( t \right) \right| \right)} \nonumber\\
&\le \bar{\theta}\mathbb{L}_{\pi _i}e_{i,h}c_{i,m}\sum_{k=1}^{n_i}{\left( \left| z_{i,k}\left( t \right) -z_{i,k}\left( t-\tau _h \right) \right|+\left| z_{i,k}\left( t \right) \right| \right)} \nonumber\\
&\le \bar{\theta}\mathbb{L}_{\pi _i}c_{i,m}e_{i,h}\left( e_{i,h}+z_i \right)
\end{align}
Substitute~(\ref{43}) and~(\ref{44}) into~(\ref{41}), it yields
\begin{align}\label{45}
&\left( z_{i,n_i}\pi \left( x_i \right) -z_{i,n_i}\left( t-\tau _h \right) \pi _i\left( x_i\left( t-\tau _h \right) \right) \right)^{T}\theta _{i}^{*} \nonumber\\
&\le z_i\bar{\theta}\mathbb{L}_{\pi _i}c_{i,m}e_{i,h}+\bar{\theta}\mathbb{L}_{\pi _i}c_{i,m}e_{i,h}\left( e_{i,h}+z_i \right)
\end{align}
In the second term~(\ref{40}), from~(\ref{32}) and~(\ref{42}), one obtains
\begin{align}\label{46}
&\left( z_{i,n_i}\left( t-\tau _h \right) -z_{i,n_i} \right) \hat{\theta}_{i}^{T}\left( t-\tau _h \right) \pi _i\left( x_i\left( t-\tau _h \right) \right)\nonumber\\
&\le \bar{\theta}\mathbb{L}_{\pi _i}e_{i,h}\sum_{k=1}^{n_i}{\left( \left| x_{i,k}\left( t \right) -x_{i,k}\left( t-\tau _h \right) \right|+\left| x_{i,k}\left( t \right) \right| \right)} \nonumber\\
&\le \bar{\theta}\mathbb{L}_{\pi _i}e_{i,h}c_{i,m}\sum_{k=1}^{n_i}{\left( \left| z_{i,k}\left( t \right) -z_{i,k}\left( t-\tau _h \right) \right|+\left| z_{i,k}\left( t \right) \right| \right)} \nonumber\\
&\le \bar{\theta}\mathbb{L}_{\pi _i}c_{i,m}e_{i,h}\left( e_{i,h}+z_i \right)
\end{align}

Therefore, for~(\ref{30}), by deriving the bound of the above two terms, substitute~(\ref{45}) and~(\ref{46}) into~(\ref{38}), the boundary obtained as
\begin{align}\label{47}
&z_{i,n_i}\left( \bar{u}_{i,2}\left( t-\tau _h \right) +\pi^{T} \left( x_i \right)\theta _{i}^{*} \right) -\tilde{\theta}_{i}^{T}\varGamma _{i}^{-1}\varphi _i\left( t-\tau _h \right) \nonumber\\
&\le z_i\bar{\theta}\mathbb{L}_{\pi _i}c_{i,m}e_{i,h}+2\bar{\theta}\mathbb{L}_{\pi _i}c_{i,m}e_{i,h}\left( e_{i,h}+z_i \right)
\end{align}
For~(\ref{31}), from~(\ref{13}) and~(\ref{17}), one knows
\begin{align}\label{48}
&z_{i,n_i}\left( \bar{u}_{i,3}\left( t-\tau _h \right) +\lambda _{i}^{*}\varOmega \left( x_i \right) \right) -\tilde{\lambda}_i\zeta _{i}^{-1}\phi _i\left( t-\tau _h \right)
\nonumber\\
&=z_{i,n_i}\left( \lambda _{i}^{*}\varOmega \left( x_i \right) -\hat{\lambda}_{i}^{}\left( t-\tau _h \right) \varOmega _i\left( x_i\left( t-\tau _h \right) \right) \right)
\nonumber\\
&\hspace{1em}-\left( \lambda _{i}^{*}-\hat{\lambda}_i\left( t-\tau _h \right) \right) z_{i,n_i}\left( t-\tau _h \right) \varOmega _i\left( x_i\left( t-\tau _h \right) \right)
\nonumber\\
&=\lambda _{i}^{*}\left( z_{i,n_i}\varOmega \left( x_i \right) -z_{i,n_i}\left( t-\tau _h \right) \varOmega _i\left( x_i\left( t-\tau _h \right) \right) \right)
\nonumber\\
&\hspace{1em}+\left( z_{i,n_i}\left( t-\tau _h \right) -z_{i,n_i} \right) \hat{\lambda}_{i}^{}\left( t-\tau _h \right) \varOmega _i\left( x_i\left( t-\tau _h \right) \right)
\end{align}
Then divide the result of~(\ref{48}) into two terms,
\begin{equation}\label{49}
\lambda _{i}^{*}\left( z_{i,n_i}\varOmega \left( x_i \right) -z_{i,n_i}\left( t-\tau _h \right) \varOmega _i\left( x_i\left( t-\tau _h \right) \right) \right)
\end{equation}
and
\begin{equation}\label{50}
\left( z_{i,n_i}\left( t-\tau _h \right) -z_{i,n_i} \right) \hat{\lambda}_{i}^{}\left( t-\tau _h \right) \varOmega _i\left( x_i\left( t-\tau _h \right) \right)
\end{equation}
Derive the bounds for~(\ref{49}) and~(\ref{50}), respectively.

In the first term~(\ref{49}), add and subtract the same term $z_{i,n_i}\varOmega\left( x_i\left( t-\tau _h \right) \right)
\lambda _{i}^{*}$, keep unchanged formula as
\begin{align}\label{51}
&\lambda _{i}^{*}\left( z_{i,n_i}\varOmega \left( x_i \right) -z_{i,n_i}\left( t-\tau _h \right) \varOmega _i\left( x_i\left( t-\tau _h \right) \right) \right)
\nonumber\\
&=\left(z_{i,n_i}\varOmega \left( x_i \right) -z_{i,n_i}\varOmega _i\left( x_i\left( t-\tau _h \right) \right) \right) \lambda _{i}^{*}
\nonumber\\
&\hspace{1em}+\left( z_{i,n_i}-z_{i,n_i}\left( t-\tau _h \right) \right) \varOmega _i\left( x_i\left( t-\tau _h \right) \right) \lambda _{i}^{*}
\end{align}
According to Assumption 4,~(\ref{19}),~(\ref{24}) and~(\ref{32}), one knows
\begin{align}\label{52}
\left| \varOmega _i -\varOmega _i\left( x_i\left( t-\tau _h \right) \right) \right|
&\le \mathbb{L}_{\varOmega_i}\sum_{k=1}^{n_i}{\left| x_{i,k} -x_{i,k}\left( t-\tau _h \right) \right|}\nonumber\\
 &\le \mathbb{L}_{\varOmega_i}\sum_{k=1}^{n_i}{\bar{c}_{i,k}\left| z_{i,k} -z_{i,k}\left( t-\tau _h \right) \right|}\nonumber\\
 &\le \mathbb{L}_{\varOmega_i}c_{i,m}e_{i,h}
\end{align}
From~(\ref{32}) and~(\ref{52}), one obtains
\begin{equation}\label{53}
z_{i,n_i}\left(\varOmega \left( x_i \right) -\varOmega _i\left( x_i\left( t-\tau _h \right) \right) \right) \lambda _{i}^{*}\le z_i\bar{\lambda}\mathbb{L}_{\varOmega _i}c_{i,m}e_{i,h}
\end{equation}
and
\begin{align}\label{54}
&\left( z_{i,n_i}-z_{i,n_i}\left( t-\tau _h \right) \right) \varOmega _i\left( x_i\left( t-\tau _h \right) \right) \lambda _{i}^{*}\nonumber\\
&\le \bar{\lambda}\mathbb{L}_{\varOmega _i}e_{i,h}\sum_{k=1}^{n_i}{\left( \left| x_{i,k}\left( t \right) -x_{i,k}\left( t-\tau _h \right) \right|+\left| x_{i,k}\left( t \right) \right| \right)} \nonumber\\
&\le \bar{\lambda}\mathbb{L}_{\varOmega _i}e_{i,h}c_{i,m}\sum_{k=1}^{n_i}{\left( \left| z_{i,k}\left( t \right) -z_{i,k}\left( t-\tau _h \right) \right|+\left| z_{i,k}\left( t \right) \right| \right)} \nonumber\\
&\le \bar{\lambda}\mathbb{L}_{\varOmega _i}c_{i,m}e_{i,h}\left( e_{i,h}+z_i \right)
\end{align}
Substitute~(\ref{53}) and~(\ref{54}) into~(\ref{51}), it yields
\begin{align}\label{55}
&\lambda _{i}^{*}\left( z_{i,n_i}\varOmega \left( x_i \right) -z_{i,n_i}\left( t-\tau _h \right) \varOmega _i\left( x_i\left( t-\tau _h \right) \right) \right) \nonumber\\
&\le z_i\bar{\lambda}\mathbb{L}_{\varOmega _i}c_{i,m}e_{i,h}+\bar{\lambda}\mathbb{L}_{\varOmega _i}c_{i,m}e_{i,h}\left( e_{i,h}+z_i \right)
\end{align}
In the second term~(\ref{50}), from~(\ref{32}) and~(\ref{52}), one obtians
\begin{align}\label{56}
&\left( z_{i,n_i}\left( t-\tau _h \right) -z_{i,n_i} \right) \hat{\lambda}_{i}^{}\left( t-\tau _h \right) \varOmega _i\left( x_i\left( t-\tau _h \right) \right)\nonumber\\
&\le \bar{\lambda}\mathbb{L}_{\varOmega _i}e_{i,h}\sum_{k=1}^{n_i}{\left( \left| x_{i,k}\left( t \right) -x_{i,k}\left( t-\tau _h \right) \right|+\left| x_{i,k}\left( t \right) \right| \right)} \nonumber\\
&\le \bar{\lambda}\mathbb{L}_{\varOmega _i}e_{i,h}c_{i,m}\sum_{k=1}^{n_i}{\left( \left| z_{i,k}\left( t \right) -z_{i,k}\left( t-\tau _h \right) \right|+\left| z_{i,k}\left( t \right) \right| \right)} \nonumber\\
&\le \bar{\lambda}\mathbb{L}_{\varOmega _i}c_{i,m}e_{i,h}\left( e_{i,h}+z_i \right)
\end{align}
For~(\ref{31}), by deriving the bound of the above two terms, substitute~(\ref{55}) and~(\ref{56}) into~(\ref{48}), the bound obtained as
\begin{align}\label{57}
&z_{i,n_i}\left( \bar{u}_{i,2}\left( t-\tau _h \right) +\lambda _{i}^{*}\varOmega \left( x_i \right) \right) -
\tilde{\lambda}_i\zeta _{i}^{-1}\phi _i\left( t-\tau _h \right) \nonumber\\
&\le z_i\bar{\lambda}\mathbb{L}_{\varOmega _i}c_{i,m}e_{i,h}+2\bar{\lambda}\mathbb{L}_{\varOmega _i}c_{i,m}e_{i,h}\left( e_{i,h}+z_i \right)
\end{align}

By analyzing the discrepancy of each part of the Lyapunov function, substitute~(\ref{37}),~(\ref{47}) and~(\ref{57}) into~(\ref{26}), for $i\in\mathcal{B}_1$,$\mathcal{G}_i\neq \varnothing$, one knows
\begin{align}\label{58}
 \dot{V}_{i,n_i}\le &-\sum_{k=1}^{n_i}{\gamma z_{i,k}^{2}}+z_{i,n_i}e_{i,h}\left( b_{i,m}+\gamma+\mathbb{L}_{\alpha_i}c_{i,m} \right)\nonumber\\
&+z_i\bar{\theta}\mathbb{L}_{\pi _i}c_{i,m}e_{i,h}+2\bar{\theta}\mathbb{L}_{\pi _i}c_{i,m}e_{i,h}\left( e_{i,h}+z_i \right)\nonumber\\
 &+ z_i\bar{\lambda}\mathbb{L}_{\varOmega _i}c_{i,m}e_{i,h}+2\bar{\lambda}\mathbb{L}_{\varOmega _i}c_{i,m}e_{i,h}\left( e_{i,h}+z_i \right)\nonumber\\
 &+z_{i,n_i}\mathbb{L}_{\beta_i}\sum_{j\in \mathcal{G}_i}^{}{c_{j,m}e_{j,h}}
 \end{align}

For $i\in\mathcal{B}_2$, the state of the interconnection function is under normal, that mean $\mathcal{G}_i=\varnothing$. According to the derivation process for~(\ref{58}), the Lyapunov function becomes
\begin{align}\label{59}
 \dot{V}_{i,n_i}\le &-\sum_{k=1}^{n_i}{\gamma z_{i,k}^{2}}+z_{i,n_i}e_{i,h}\left( b_{i,m}+\gamma+\mathbb{L}_{\alpha_i}c_{i,m} \right)\nonumber\\
&+z_i\bar{\theta}\mathbb{L}_{\pi _i}c_{i,m}e_{i,h}+2\bar{\theta}\mathbb{L}_{\pi _i}c_{i,m}e_{i,h}\left( e_{i,h}+z_i \right)\nonumber\\
 &+ z_i\bar{\lambda}\mathbb{L}_{\varOmega _i}c_{i,m}e_{i,h}+2\bar{\lambda}\mathbb{L}_{\varOmega _i}c_{i,m}e_{i,h}\left( e_{i,h}+z_i \right)
 \end{align}

\textbf{Part\ 2:} Normal subsystem

Then let's consider the case $ i\in \mathcal{H} $, which means that the $ i $-th subsystem is normal. Figs. \ref{Fig2} and \ref{Fig3} reveal that the subsystem not subjected to the network anomaly will not experience changes in its own state. However, due to information interaction between subsystems, the interconnection function $\beta _i$ is influenced by neighboring states and may turn to be $\beta _i\left( \bar{x}_i\left( t-\tau _h \right) \right)$ for control design. Consequently, for $ i\in \mathcal{H}_1 $, the interconnection function discrepancy due to network anomaly for the distributed controller is induced. Then, the derivative of the Lyapunov function for the $i$-th subsystem is
\begin{align}\label{60}
 \dot{V}_{i,n_i}&=-\sum_{k=1}^{n_i}{\gamma z_{i,k}^{2}}+z_{i,n_i}\left( \beta _i\left( \bar{x}_i \right) -\beta _i\left( \bar{x}_i\left( t-\tau _h \right) \right) \right)
\end{align}
Taking~(\ref{28}) into~(\ref{60}), one yields
\begin{align}\label{61}
\dot{V}_{i,n_i}&\le -\sum_{k=1}^{n_i}{\gamma z_{i,k}^{2}}+z_{i,n_i}\mathbb{L}_{\beta_i}\sum_{j\in \mathcal{G}_i}^{}{c_{j,m}e_{j,h}}
\end{align}
For $ i\in \mathcal{H}_2 $, the subsystem itself and the interconnection parts are not affected by the network anomaly, then on the base of~(\ref{17}) it easy to prove that $ \dot{V}_{i,n_i}\le -\sum_{k=1}^{n_i}{\gamma z_{i,k}^{2}} $.

\textbf{Part\ 3:} Stability analysis of the entire interconnected system

For the interconnected system, define the sum of Lyapunov functions as
\begin{equation}\label{62}
  V_M=\frac{1}{2}\sum_{i=1}^M{\sum_{k=1}^{n_i}{z_{i,k}^{2}}+\frac{1}{2}\sum_{i=1}^M\left(\tilde{\theta}_{i}^{T}\varGamma _{i}^{-1}\tilde{\theta}_i+\frac{1}{\zeta _i}\tilde{\lambda}_{i}^{2} \right)}
\end{equation}

According to the received state of~Fig.\ref{Fig3}, different subsystems for the distributed controllers of the nonlinear interconnected system under multiple network anomalies, the Lyapunov function of the interconnected system is given by
\begin{align}\label{63}
\dot{V}_M\le&-\sum_{i=1}^M{\sum_{k=1}^{n_i}{\gamma z_{i,k}^{2}}}+\sum_{i\in \mathcal{B}}^{}{z_{i,n_i}\left( \bar{u}_i\left( t-\tau _h \right) -\bar{u}_{i,1} \right)} \nonumber\\
 &+\sum_{i\in \mathcal{B}}^{}z_{i,n_i}\pi^{T} \left( x_i \right)\theta _{i}^{*} -\tilde{\theta}_{i}^{T}\varGamma _{i}^{-1}\varphi _i\left( t-\tau _h \right) \nonumber\\
  &+\sum_{i\in \mathcal{B}}^{}z_{i,n_i}\lambda _{i}^{*}\varOmega \left( x_i \right) -\tilde{\lambda}_i\zeta _{i}^{-1}\phi _i\left( t-\tau _h \right)\nonumber\\
  &+\sum_{i\in \mathcal{H}_1}^{}{z_{i,n_i}\left( \beta _i\left( \bar{x}_i \right) -\beta _i\left( \bar{x}_i\left( t-\tau _h \right) \right) \right)}
\end{align}
From~(\ref{58}),~(\ref{59}) and~(\ref{61}), one knows
\begin{align}\label{64}
\dot{V}_M\le&-\sum_{i=1}^M{\sum_{k=1}^{n_i}{\gamma z_{i,k}^{2}}}+\sum_{i\in \mathcal{H}_1}^{}\left({z_{i,n_i}\mathbb{L}_{\beta _i}\sum_{j\in \mathcal{G}_i}^{}{c}_{j,m}e_{j,h}}\right)
 \nonumber\\
 &+\sum_{i\in \mathcal{B}_1}^{}\left({z_{i,n_i}\mathbb{L}_{\beta _i}\sum_{j\in \mathcal{G}_i}^{}{c}_{j,m}e_{j,h}}\right)\nonumber\\
&+\sum_{i\in \mathcal{B}}^{}c_{i,m}e_{i,h}\left( \bar{\theta}\mathbb{L}_{\pi _i}+\bar{\lambda}\mathbb{L}_{\varOmega _i} \right) \left( 3z_i+2e_{i,h} \right) \nonumber\\
&+\sum_{i\in \mathcal{B}}^{}{z}_{i,n_i}e_{i,h}\left( b_{i,m}+\gamma +\mathbb{L}_{\alpha _i}c_{i,m} \right)
\end{align}

Since the specific subsystems to be anomalous remain unknown, the tolerance and boundary conditions of the interconnected system are considered by investigating the dynamics of $ \frac{e_h}{\sum_{i=1}^M{z_i}} $.

Taking the time derivative of $  \frac{e_h}{\sum_{i=1}^M{z_i}}   $ gives
\begin{equation}\label{65}
\frac{d}{dt}\frac{e_h}{\sum_{i=1}^M{z_i}}=\frac{\dot{e}_h}{\sum_{i=1}^M{z_i}}-\frac{e_h\sum_{i=1}^M\dot{z_i}}{\left(\sum_{i=1}^M{z_i}\right)^{2}}
\end{equation}
In the following, we will analyze the bounds of $\dot{e}_h$ and $\sum_{i=1}^M\dot{z_i}$, respectively.

First, we analyze the bound of $\dot{e}_h$. Based on~(\ref{32}), when $ t\in \left[ t_h,t_h+t_{d,h} \right)  $, one knows
\begin{align}\label{66}
\dot{e}_{i,h}&=\sum_{k=1}^{n_i}{sgn \left( z_{i,k} -z_{i,k}\left( t-\tau _h \right) \right) \left( \dot{z}_{i,k}-\dot{z}_{i,k}\left( t-\tau _h \right) \right)}\nonumber\\
&\le \sum_{k=1}^{n_i}{\left| \dot{z}_{i,k} -\dot{z}_{i,k}\left( t-\tau _h \right) \right|}
\end{align}
Thus, from~(\ref{32}), we know $e_h=\sum_{i=1}^M{\sum_{k=1}^{n_i}{e_{i,h}}}$. According to~(\ref{66}), it can obtain
\begin{equation}\label{67}
 \dot{e}_{h}\le \sum_{i=1}^M\sum_{k=1}^{n_i}{\left| \dot{z}_{i,k} -\dot{z}_{i,k}\left( t-\tau _h \right) \right|}
\end{equation}
From~(\ref{18}), one obtains
\begin{align}\label{68}
\left| \dot{z}_{i,1}-\dot{z}_{i,1}\left( t-\tau _h \right)\right|\le  &\ \gamma \left| z_{i,1}-z_{i,1}\left( t-\tau _h \right)  \right|\nonumber\\ 	
 &+ \left|  z_{i,2}-z_{i,2}\left( t-\tau _h \right)\right|
\end{align}
also for $2\le j \le n_i-1$,
\begin{align}\label{69}
\left| \dot{z}_{i,j}-\dot{z}_{i,j}\left( t-\tau _h \right)  \right| \le \ & \gamma \left|  z_{i,j}-z_{i,j}\left( t-\tau _h \right)  \right|\nonumber\\ 	
&+\left|  z_{i,j-1  }-z_{i, j-1  }\left( t-\tau _h \right)  \right| \nonumber\\
&+\left|  z_{i, j+1 }-z_{i, j+1  }\left( t-\tau _h \right)  \right|
\end{align}
From~(\ref{18}),~(\ref{42}), and~(\ref{52}), one yields
\begin{align}\label{70}
\left|  \dot{z}_{i,n_i}-\dot{z}_{i,n_i}\left( t-\tau _h \right) \right| \le &\ \gamma \left|  z_{i,n_i}-z_{i,n_i}\left( t-\tau _h \right)  \right|\nonumber\\ 	
&+\left|  z_{i, n_i-1  }-z_{i, n_i-1  }\left( t-\tau _h \right)  \right| \nonumber\\
&+2c_{i,m}e_{i,h}\left( \bar{\theta}\mathbb{L}_{\pi _i}+\bar{\lambda}\mathbb{L} \right)
\end{align}
Therefore, combining~(\ref{68}), ~(\ref{69}) and~(\ref{70}), one knows
\begin{align}\label{71}
\dot{e}_{i,h}&\le\sum_{k=1}^{n_i}{\left| \dot{z}_{i,k}\left( t \right) -\dot{z}_{i,k}\left( t-\tau _h \right) \right|}\nonumber \\
&\le e_{i,h}\left( \gamma +2 +2c_{i,m}\left( \bar{\theta}\mathbb{L}_{\pi_i}+\bar{\lambda}\mathbb{L}_{\varOmega_i} \right)\right)
\end{align}
From~(\ref{22}) and~(\ref{32}), and taking~(\ref{71}) into~(\ref{67}), one can obtain
\begin{align}\label{72}
\dot{e}_h\le e_h\left( \gamma +2+2c_m\left( \bar{\theta}\mathbb{L}_{\pi}+\bar{\lambda}\mathbb{L}_{\varOmega}\right)\right)
\end{align}
Thus, combining~(\ref{65}) and~(\ref{72}), take the time-derivative of $ \frac{e_h}{\sum_{i=1}^M{z_i}} $ satisfies
\begin{align}\label{73}
\frac{d}{dt}\frac{e_h}{\sum_{i=1}^M{z_i}}\le& \frac{e_h}{\sum_{i=1}^M{z_i}}\left(  \gamma +2+c_m\left( \bar{\theta} \mathbb{L}_{\pi}+\bar{\lambda}\mathbb{L}_{\varOmega}\right)  \right)\nonumber\\
&+\frac{e_h}{\sum_{i=1}^M{z_i}}\frac{\sum_{i=1}^M\dot{z_i}}{\sum_{i=1}^M{z_i}}
\end{align}

Then, we analyze the bound of $\sum_{i=1}^M\dot{z_i}$.
From~(\ref{18}) and~(\ref{32}), one gets
\begin{equation}\label{74}
\left| \dot{z}_{i,1} \right|+...+\left| \dot{z}_{i, n_i-1 } \right|\le z_i\left( \gamma +2 \right)
\end{equation}
and
\begin{align}\label{75}
\dot{z}_{i,n_i}&=\alpha _i\left( x_i \right) +\bar{u}_i\left( t-\tau _h \right) +\eta _i\left( x_i,t \right) +\beta _i\left( \bar{x}_i \right) \nonumber \\
&-\sum_{k=1}^{n_i-1}{\frac{\partial \alpha _{i,n_i-1}}{\partial x_{i,k}}x_{i,k+1}}
\end{align}
Taking~(\ref{25}) into~(\ref{75}), it yields
\begin{align}\label{76}
\dot{z}_{i,n_i}\le& -\gamma z_{i,n_i}\left( t-\tau _h \right) -z_{i,n_i-1}\left( t-\tau _h \right) \nonumber \\
&+\alpha _i\left( x_i \right) -\alpha _i\left( x_i\left( t-\tau _h \right) \right)\nonumber \\
&+\beta _i\left( \bar{x}_i \right) -\beta _i\left( \bar{x}_i\left( t-\tau _h \right) \right) +b_{i,m}e_{i,h}\nonumber \\
&+\pi^{T} \left( x_i \right)\theta _{i}^{*} -\hat{\theta}_{i}^{T}\left( t-\tau _h \right) \pi \left( x_i\left( t-\tau _h \right) \right)\nonumber \\
&+\lambda _{i}^{*}\varOmega \left( x_i \right) -\hat{\lambda}_{i}^{}\left( t-\tau _h \right) \varOmega \left( x_i\left( t-\tau _h \right) \right)
\end{align}
From~(\ref{42}), it can obtain
\begin{align}\label{77}
\left(\pi _i\left( x_i \right)- \pi _i\left( x_i\left( t-\tau _h \right) \right)\right)^{T}{\theta}_{i}^{*} \le \bar{\theta}c_{i,m}\mathbb{L}_{\pi _i}e_{i,h}
\end{align}
and
\begin{align}\label{78}
&\left({\theta}_{i}^{*} -\hat{\theta}_{i}^{}\left( t-\tau _h \right) \right)^T\pi _i\left( x_i \left( t-\tau _h \right)\right) \nonumber \\
&\le 2c_{i,m} \bar{\theta}\mathbb{L}_{\pi _i}\left(e_{i,h}+z_{i}\right)
\end{align}
Combining~(\ref{77}) and~(\ref{78}), one gets
\begin{align}\label{79}
&\pi^{T} \left( x_i \right)\theta _{i}^{*}-\hat{\theta}_{i}^{T}\left( t-\tau _h \right) \pi \left( x_i\left( t-\tau _h \right) \right)\nonumber\\
&\le c_{i,m} \bar{\theta}\mathbb{L}_{\pi _i}\left(3e_{i,h}+2z_i \right)
\end{align}
Similarly in the same way as following the derivation process for~(\ref{79}), one can obtain
\begin{align}\label{80}
&\lambda _{i}^{*}\varOmega  \left( x_i \right) -\hat{\lambda}_{i}^{}\left( t-\tau _h \right) \varOmega  \left( x_i\left( t-\tau _h \right) \right)\nonumber\\
&\le c_{i,m} \bar{\lambda}\mathbb{L}_{\varOmega _i}\left(3e_{i,h}+2z_i \right)
\end{align}
Taking~(\ref{34}),~(\ref{35}),~(\ref{79}) and~(\ref{80}) into~(\ref{76}), it yields
\begin{align}\label{81}
\left| \dot{z}_{i,n_i} \right|\le&\hspace{0.5em} c_{i,m}\left( \bar{\theta}\mathbb{L}_{\pi}+\bar{\lambda}_i\mathbb{L}_{\varOmega _i} \right) \left( 3e_{i,h}+2z_i \right) \mspace{18mu}
\nonumber\\
&+c_{i,m}e_{i,h}\mathbb{L}_{\alpha _i}+e_{i,h}\left( \gamma +b_{i,m} \right) \mspace{18mu}
\nonumber\\
&+\gamma z_i+\mathbb{L}_{\beta _i}\sum_{j\in \mathcal{G}_i}^{}{c_{j,m}e_{j,h}}
\end{align}
Merging~(\ref{74}) and~(\ref{81}), one knows
\begin{align}\label{82}
\dot{z}_i\le&\hspace{0.5em}c_{i,m}\left( \bar{\theta}\mathbb{L}_{\pi}+\bar{\lambda}_i\mathbb{L}_{\varOmega _i} \right) \left( 3e_{i,h}+2z_i \right) \nonumber \\
&+c_{i,m}e_{i,h}\mathbb{L}_{\alpha _i}+e_{i,h}\left( \gamma +b_{i,m} \right)\nonumber \\
&+2z_i\left( \gamma +1 \right)+\mathbb{L}_{\beta _i}\sum_{j\in \mathcal{G}_i}^{}{c_{j,m}e_{j,h}}
\end{align}
For the interconnected system, from~(\ref{82}), one gets
\begin{align}\label{83}
\sum_{i=1}^M\dot{z_i}\le&\  e_hc_m\left( \mathbb{L}_{\alpha}+3\bar\theta \mathbb{L}_{\pi}+3\bar\lambda \mathbb{L}_{\varOmega}+M\mathbb{L}_{\beta} \right) \nonumber \\
&+2\sum_{i=1}^M{z_i}\left( \gamma +1+c_m\left(\bar\theta\mathbb{L}_{\pi}+\bar\lambda\mathbb{L}_{\varOmega}\right) \right)\nonumber \\
&+e_h\left( \gamma +b_m \right)
\end{align}
Then, taking~(\ref{83}) into~(\ref{73}), one can obtain
\begin{equation}\label{84}
\frac{d}{dt}\frac{e_h}{\sum_{i=1}^M{z_i}}\le \left( 1+\frac{e_h}{\sum_{i=1}^M{z_i}} \right) \left( w_1+w_2\frac{e_h}{\sum_{i=1}^M{z_i}} \right)
\end{equation}
where $ w_1$ and $w_2$ are given in~(\ref{21}).
For $ t\in \left[ t_h,t_h+t_{d,h} \right)  $, by solving the differential inequality~(\ref{84}), it yields
\begin{equation}\label{85}
   \frac{e_h}{\sum_{i=1}^M{z_i}}\le \frac{e^{\left( w_1-w_2 \right) t_{d,h}}-1}{1-\frac{w_2}{w_1}e^{\left( w_1-w_2 \right) t_{d,h}}}
\end{equation}
By solving $  e^{\left( w_1-w_2 \right) t_{d,h}}-1\le w_3\left( 1-\frac{w_2}{w_1}e^{\left( w_1-w_2 \right) t_{d,h}} \right)    $, where $w_3$ is given in~(\ref{21}), it can obtain that
\begin{equation}\label{86}
   t_{d,h}\le \frac{1}{w_1-w_2}ln\frac{w_1w_3+w_1}{w_2w_3+w_1}
\end{equation}
For $ t\in \left[ t_h,t_h+t_{d,h} \right)  $, from~(\ref{85}), one gets
\begin{equation}\label{87}
e_h \le \sum_{i=1}^M{w_3 z_i}
\end{equation}
From~(\ref{64}), we consider the lower bound of the interconnected system, it can get
\begin{align}\label{88}
\dot{V}_M\le&-\sum_{i=1}^M{\sum_{k=1}^{n_i}{\gamma z_{i,k}^{2}}}+2c_m\left( \bar{\lambda}\mathbb{L}_{\varOmega}+\bar{\theta}\mathbb{L}_{\pi} \right) e_{h}^{2}\nonumber\\
&+\left( c_m\left(\mathbb{L}_{\alpha}+3\bar{\lambda}\mathbb{L}_{\varOmega}+3\bar{\theta}\mathbb{L}_{\pi}+M\mathbb{L}_{\beta} \right)  \right) e_h\sum_{i=1}^M{z_i}\nonumber\\
&+\left( \gamma +b_m \right) e_h\sum_{i=1}^M{z_i}
\end{align}
Replace the parameter in~(\ref{88}) with~(\ref{21}), it yields
\begin{align}\label{89}
 \dot{V}_M\le &-\sum_{i=1}^M{\sum_{k=1}^{n_i}{\gamma z_{i,k}^{2}}}+w _2e_h\sum_{i=1}^M{z_{i}}\nonumber\\
 &+2c_m\left( \bar{\lambda}\mathbb{L}_{\varOmega}+\bar{\theta}\mathbb{L}_{\pi} \right)e_{h}^{2}
\end{align}
Taking~(\ref{87}) into~(\ref{89}), it yields
\begin{align}\label{90}
\dot{V}_M&\le -\sum_{i=1}^M{\sum_{k=1}^{n_i}{\gamma z_{i,k}^{2}}}+ w _2w _3  \left( \sum_{i=1}^M{z_i} \right) ^2 \nonumber\\
&\hspace{1.1em}+\left( 2c_m\left( \bar{\lambda}\mathbb{L}_{\varOmega}+\bar{\theta}\mathbb{L}_{\pi} \right) w _{3}^{2} \right) \left( \sum_{i=1}^M{z_i} \right) ^2 \nonumber\\
&\le -\sum_{i=1}^M{\sum_{k=1}^{n_i}{\gamma z_{i,k}^{2}}}+\frac{1}{Mn_m}\left( \sum_{i=1}^M{z_i} \right) ^2 \nonumber\\
&\le -\left( \gamma -1 \right) \sum_{i=1}^M{\sum_{k=1}^{n_i}{z_{i,k}^{2}}}\hspace{2em}
\end{align}
When $ t\in \left[ t_h,t_h+t_{d,h} \right)  $, as $\gamma>1$, $\dot V_M \le 0$, from~(\ref{90}) we know, for $ h\in Z^+ $,  $ z_{i,k}$, $\tilde{\theta}_i$, $\tilde{\lambda}_i $ are bounded, and from~(\ref{17})-(\ref{19}) we know $ u_i\left( t_h+t_{d,h} \right)  $, $ \dot{\hat{\theta}}_i\left( t_h+t_{d,h} \right)  $ and $ \dot{\hat{\lambda}}_i\left( t_h+t_{d,h} \right)  $ are bounded, also $\dot z_{i,k}\left(t_h+t_{d,h}\right)$ are bounded.
Thus, we can obtain $ \dot{V}_M\left( t_h+t_{d,h} \right)  $ is also bounded. Therefore for $ V_M $ there will be no jump which occur at $ t=t_h $, which means $ V_M\left( t_{h}^{-} \right) =V_M\left( t_{h}^{+} \right)  $.

For $ t\in \left[ 0,\infty \right)  $, and $ \gamma >1 $, it holds $\dot{V}_M\le 0$. Thus, all signals in $ V_M $, including $ z_{i,k},\tilde{\theta}_i,\tilde{\lambda}_i $ are bounded, therefore from~(\ref{18}), we know that $\dot z_{i,k}$ are bounded, $ \dot{V}_M\left( t \right)  $ is also bounded. Therefore $ V_M\left( t_{h}^{-} \right) =V_M\left( t_{h}^{+} \right)  $ and $ V_M\left( t_h+t_{d,h}^{+} \right) \le V_M\left( t_{h}^{+} \right)  $. With LaSalle-Yoshizawa Theorem, the signal are continuous and satisfy $\underset{t\rightarrow \infty}{\lim}z_{i,k}=0 $.

Therefore, under network anomalies, the uncertain nonlinear interconnected system is asymptotically stable when the time duration of each anomaly satisfied $t_{d,h}\le \frac{1}{w_1-w_2}ln\frac{w_1w_3+w_1}{w_2w_3+w_1}$.\ $\square$

In Theorem 1, quantitative resilience conditions such that the system can remain stable against network anomalies without collapsing is given. The resilience condition required for the stability of the system are derived by analyzing the Lyapunov function, where the obtained quantitative resilience condition is a function of system parameters.

\begin{remark}
\rm{In practice, the duration of each anomaly is not the same. However, for the sake of simplicity, we consider the lowest tolerance, i.e., the longest time anomalies can be tolerated. Since the topology of the anomalous subsystems remains unknown and alterable, the resilience condition is more or less conservative. The conservativeness of the resilience condition can be released when the topology of the anomalous subsystems are known. Also the technique used in the proof relies on Lipschitz continuity and must be conservative for nonlinear systems in general\cite{a29}.}
\end{remark}
\subsection{Resting time between two consecutive anomalies}
As the duration of anomaly may be conservative, a complementary resilience condition for the uncertain nonlinear interconnected system is given, where the resting time instead of the lasting time of network anomalies are analyzed. The major result is summarized by the following theorem.

\begin{theorem}
Under Assumptions 1-4 and $\gamma>1$, the uncertain nonlinear interconnected system given by~(\ref{1}) and controlled by~(\ref{17}) for any bounded initial condition are bounded, if the resting time of each network anomaly satisfies
\begin{equation}\label{91}
t_{r,h}\ge \frac{\frac{1}{2}\left( \kappa e^{\rho _1t_{d,h}}-\frac{\rho _2}{\rho _1} \right) ^2-\sum_{i=1}^M{s_i}}{2\gamma \sum_{i=1}^M{\vartheta _i}}
\end{equation}
where $\kappa =\sum_{i=1}^M{\sqrt{2n_is_i}}+\frac{\rho _2}{\rho _1}$, and
\begin{equation}\label{92}
 \left\{ \begin{array}{l} 	s_i=\max \left\{ \frac{1}{2}\sum_{k=1}^{n_i}{z_{i,k}\left( 0 \right) ^2}, \vartheta _i \right\} \\
 \rho _1=c_m\left( \mathbb{L}_{\alpha}+4\bar\theta\mathbb{L}_{\pi}+4\bar\lambda \mathbb{L}_{\varOmega}+M\mathbb{L}_{\beta} \right)+3\gamma\\  	
\hspace{2.3em}+b_m+2\\ 	
\rho _2=c_m\sum_{i=1}^M{\sqrt{2n_is_i}}\left( \mathbb{L}_{\alpha}+3\bar\theta \mathbb{L}_{\pi}+3\bar\lambda \mathbb{L}_{\varOmega}\right)\\ 	
  \hspace{2.3em}+\sum_{i=1}^M{\sqrt{2n_is_i}}\left( \gamma +b_m +M\mathbb{L}_{\beta}c_m\right)\\  \end{array} \right.
\end{equation}
where $\vartheta _i $ is a positive constant.
\end{theorem}
~\\
\textbf{Proof}: By defining a compact set, the system state can remain in the compact set after the network anomaly, and then the entire interconnected system can remain bounded. After an anomaly on the system, the system state will deviate from the compact set, and the resting time is required to pull the system state back to the compact set before the next anomaly. The objective is to find the resilient condition for which the minimum resting time is such that the system state returns to the compact set and the interconnected system is bounded.

Define a compact set $\mathcal{S}  $ as
\begin{equation}\label{93}
 \mathcal{S}=\left\{ \left( z_{i,k},\tilde{\theta}_i,\tilde{\lambda}_i \right) :V_M\le s \right\} \
\end{equation}
where $ k=\left\{ 1,...,n_i \right\}  $, $ i=\left\{ 1,...,M \right\}  $, $ s=\sum_{i=1}^M{s_i}+s_{\theta}+s_{\lambda} $,
\begin{align}\label{94}
\left\{ \begin{array}{l}
	s_i=\max \left\{ \frac{1}{2}\sum_{k=1}^{n_i}{z_{i,k}\left( 0 \right) ^2},\vartheta _i \right\}\\
	s_{\theta}=2M\bar\theta^{2}\\
	s_{\lambda}=2M\bar\lambda^{2}\\
\end{array} \right.
\end{align}
and the $\vartheta _i \left(i=1,...,M\right) $ is a positive constant to determine the size of the compact set $ \mathcal{S} $.
We aim to ensure that all system states remain within a compact set, even after a network anomaly, with the system state eventually returning to the compact set. We provide a detailed analysis of the state boundaries across different time intervals: before the anomaly, during the anomaly, and during the resting time following the anomaly.

Obviously, before the anomaly, $ \left( z_{i,k},\tilde{\theta}_i,\tilde{\lambda}_i \right) \in \mathcal{S} $ for $ t\in \left[ 0,t_1  \right)  $. Then, it has
\begin{equation}\label{95}
  \sum_{i=1}^M{z_i}\le \sum_{i=1}^M{\sqrt{2n_is_i}}
\end{equation}
We assume that $ \left( z_{i,k},\tilde{\theta}_i,\tilde{\lambda}_i \right) \in \mathcal{S} $ for $ t\in \left[ 0,t_h \right)  $.
From~(\ref{32}) and~(\ref{95}), one can obtain
\begin{align}\label{96}
 e_{i,h}&=\sum_{k=1}^{n_i}{\left|  z_{i,k}\left( t \right) -z_{i,k}\left( t-\tau _h \right)  \right|}\nonumber \\
 &\le \left| z_i \right|+\left| z_i\left( t-\tau _h \right) \right|\nonumber \\
 &\le z_i+\sqrt{2n_is_i}
\end{align}
Also, one gets
\begin{align}\label{97}
e_h\le \sum_{i=1}^M{\left( z_i+\sqrt{2n_is_i} \right)}
\end{align}
From~(\ref{83}) and~(\ref{97}), one knows
\begin{align}\label{98}
 \sum_{i=1}^M\dot{z_i}\le \rho _1\sum_{i=1}^M{z_i}+\rho _2
\end{align}
where $\rho _1$ and $\rho _2$ are given in~(\ref{92}).

During the anomaly, for $ t\in \left[ t_h,t_h+t_{d,h} \right)  $, by solving~(\ref{98}), it yields
\begin{equation}\label{99}
  \sum_{i=1}^M{z_i}\le \left( \sum_{i=1}^M{\sqrt{2n_is_i}}+\frac{\rho _2}{\rho _1} \right) e^{\rho _1t_{d,h}}-\frac{\rho _2}{\rho _1}
\end{equation}
Therefore, it can be concluded that at the moment the anomaly ends
\begin{align}\label{100}
 V_M\left( t_h+t_{d,h} \right) \le& \frac{1}{2}\left( \kappa e^{\rho _1t_{d,h}}-\frac{\rho _2}{\rho _1} \right) ^2 +2M\bar\theta ^{2}\nonumber\\
&+2M\bar\lambda^{2}
\end{align}
where $\kappa$ is given in~(\ref{91}).

During the resting time following the anomaly, for $ t\in \left( t_h+t_{d,h},t_{h+1} \right]  $,
\begin{equation}\label{101}
 \dot{V}_M=-\sum_{i=1}^M{\sum_{k=1}^{n_i}{\gamma z_{i,k}^{2}}}
\end{equation}

However, it is necessary to determine whether the system state can return to the compact set by the end of the resting period. Therefore, we will focus on analyzing the system's behavior from the moment the anomaly ends until the resting period concludes.
At the end of the network anomaly time $t=t_h+t_{d,h}$, we analyze whether the system signals are in the secure compact set $\mathcal{S}$, and we divide it into two cases.

\textbf{Case\ 1:} If $\left( z_{i,k},\tilde{\theta}_i,\tilde{\lambda}_i \right)\notin \mathcal{S} $ at $ t=t_h+t_{d,h} $, then we have $ \sum_{k=1}^{n_i}{z_{i,k}^{2}}\ge 2\vartheta _i $, which means
\begin{equation}\label{102}
\dot{V}_M\le -2\gamma \sum_{i=1}^M{\vartheta _i}
\end{equation}
For $ t\in \left( t_h+t_{d,h},t_{h+1} \right]$, by evaluating the definite integral over this period of time, one knows
\begin{equation}\label{103}
\int_{t_h+t_{d,h}}^{t_{h+1}}{\dot{V}_M}\le \int_{t_h+t_{d,h}}^{t_{h+1}}{-2\gamma \sum_{i=1}^M{\vartheta _i}}
\end{equation}
From~(\ref{103}), it can obtain
\begin{equation}\label{104}
V_M\left( t_{h+1} \right) \le -2\gamma \sum_{i=1}^M{\vartheta _i}t_{r,h}+V_M\left( t_h+t_{d,h} \right)
\end{equation}
Substitute~(\ref{100}) into~(\ref{104}), one yields
\begin{align}\label{105}
V_M\left( t_{h+1} \right) \le &\hspace{0.5em}\frac{1}{2}\left( \kappa e^{\rho _1t_{d,h}}-\frac{\rho _2}{\rho _1} \right) ^2\nonumber\\
&-2\gamma \sum_{i=1}^M{\vartheta _i}t_{r,h}+2M\bar{\theta}^2+2M\bar{\lambda}^2
\end{align}
We expect the system state to return to the compact set at the time $t=t_{h+1}$. If the resting time of the $h$-th anomaly $t_{r,h}$ satisfies~(\ref{91}), the $V_M\left( t_{h+1} \right) \le s$ hold.
The objective is $V_M\left( t_{h+1} \right) \le s$, and $ \left( z_{i,k},\tilde{\theta}_i,\tilde{\lambda}_i \right) \in \mathcal{S} $ at $ t=t_{h+1} $ can be derived.

This means that the resting time satisfy~(\ref{91}), the system state can return to the compact set $\mathcal{S}$ before the start of the next anomaly.

\textbf{Case\ 2:} If $\left( z_{i,k},\tilde{\theta}_i,\tilde{\lambda}_i \right) \in \mathcal{S} $ at $ t=t_h+t_{d,h} $, then the conclusion holds automatically for $ t=t_{h+1} $.

Therefore, we can obtain that $ \left( z_{i,k},\tilde{\theta}_i,\tilde{\lambda}_i \right) \in \mathcal{S} $ for $ t\in \left[ 0,t_{h+1} \right)  $, and all signals of the interconnected system are bounded. It completes the proof.\ $\square$

The proof of the above two theorems illustrate that the interconnected system under the proposed distributed adaptive controller~(\ref{17}) remain bounded even in the presence of network anomalies if the quantitative resilient conditions hold. If the duration of each anomaly is less than the obtained boundary constant~(\ref{20}) or the resting time between two consecutive anomalies is greater than the obtained boundary constant~(\ref{91}), it can be derived that all signals of the interconnected system are bounded.
\begin{remark}
\rm{The design of the controller is independent of the initial value of the state, but the duration of the resting time is related to the initial value of the state $ s_i=\max \left\{ \frac{1}{2}\sum_{k=1}^{n_i}{z_{i,k}\left( 0 \right) ^2}, \vartheta _i \right\}$.  This implies that generally more retrieval time is needed for driving these states far from the origin to the compact set. In general, a larger compact set $\mathcal{S} $ allows the system state to achieve a broader range of safe operation.}
\end{remark}

\begin{remark}
\rm{In fact, the time of the anomaly is not particularly long due to the limited energy of the anomaly initiator. Moreover, the longer the network anomaly time, the higher the probability of being detected by the system. To keep the anomaly stealthy, the anomaly initiator launches a network anomaly within a certain time horizon to evade detection\cite{a22,a24}.}
\end{remark}

\begin{remark}
\rm{The quantifying of anomaly duration has been considered for DoS attack and network anomalies, respectively, e.g., \cite{a8,a11,a25}. However, a more general problem is studied in this paper: the modeling uncertainty is considered by referring to distributed approximation-based control, the interconnected system is considered with complex anomaly propagation analysis, and the overall resilience is quantified by taking into consideration of the interconnections.}
\end{remark}

\section{Simulation}\label{s5}
In this section, we give a system with four interconnected subsystems to verify the effectiveness of the proposed control strategy under network anomalies\cite{a30}. For simulation, the dynamics of the $i \text{-th} \left( i\in  \left\{ 1,2,3,4 \right\} \right) $ subsystem is described as follows:
\begin{align*}
\dot{x}_{i,1}&=x_{i,2}\\
\dot{x}_{i,2}&=-\frac{m_igl_i}{G}\sin \left( x_{i,1} \right) -\frac{M}{G}x_{i,2}+u_i\\
&\quad +\eta _i\left( x_i,t \right) +\beta _i\left( \bar{x}_i \right)
\end{align*}
with
\begin{align}
&\beta _1\left( \bar{x}_1 \right) =\sin\left( x_{2,1}\right),\hspace{0.5em}\beta _2\left( \bar{x}_2 \right) = x_{1,1}x_{1,2}  +\cos \left( 0.5e^{x_{3,1}^{}} \right),  \nonumber\\
&\beta _3\left( \bar{x}_3 \right) =x_{2,1}x_{2,2}  +\cos \left( 0.5e^{x_{4,1}^{}} \right),  \hspace{0.5em}\beta _4\left( \bar{x}_4 \right) =\sin\left(x_{3,1}\right).\nonumber
\end{align}
\begin{align}
&\eta  _1\left( {x}_1 \right) =x_{1,1}\cos\left(x_{1,2}^2\right), \hspace{0.5em}\eta _2\left( {x}_2 \right) =0.5x_{2,1}x_{2,2}, \nonumber\\
&\eta _3\left( {x}_3 \right) =x_{3,1}x_{3,2},  \hspace{0.5em}\eta  _4\left( {x}_4 \right) =x_{4,1}x_{4,2}^2.\nonumber
\end{align}
where $\beta _i$ is the interconnected function, and $\eta _i\left( x_i,t \right)$ is the modeling uncertainty.
The model parameters are $ m_1=m_3=15 $, $ m_2=m_4=10 $, $ l_1=l_2=l_4=0.5$, $ l_3=0.8$, $ g=10$, $ G=9.81$, and $ M=2$. The initial value of the states are $ \left[ x_{i,1}\left( 0 \right) \ x_{i,2}\left( 0 \right) \right] ^T=\left[ 3\,\,2 \right] ^T $. Ten Gaussian basis functions uniformly centered in the region $ \left[ -1\ 1 \right]  $ are employed \cite{a31,a32}. The initial estimate value $\hat\theta_i\left(0\right)=1$ and $\hat\lambda _i\left(0\right)=1$ are chosen. The selected design parameters are $ \zeta _i=1 $, $ \varGamma _{i}^{-1}=I $, $ \gamma_1=\gamma _4=2 $, $ \gamma _2=\gamma _3=1.5 $. For multiple anomalies performance analysis, subsystems are anomalous with $ \tau _h=0.18s $.
Define the anomaly flag function $ J\left( t \right) $:
\begin{equation*}
    J\left( t \right) =\left\{ \begin{array}{l} 	1,\ \text{anomalous}\\ 	0,\ \text{normal}\\ \end{array} \right.
\end{equation*}

We split the simulation into two scenarios, the time duration of each anomaly and resting time between two consecutive anomalies.
\subsection{Simulation of the time duration of each anomaly}
\begin{figure}[htbp]
 \centering
 \includegraphics[width=3.3in]{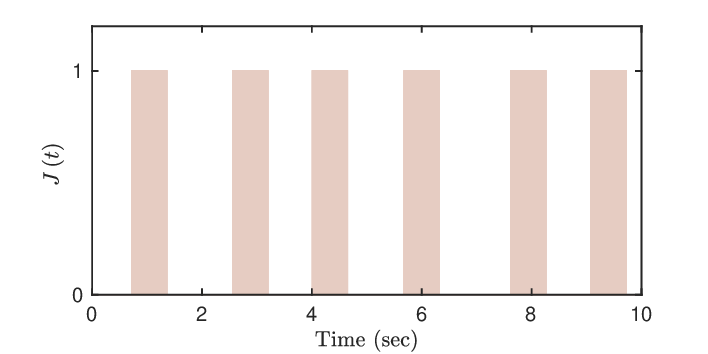}
 \caption{Anomaly time intervals. $ J\left( t \right)=1$: anomalous; $ J\left( t \right)=0$: normal.}
 \vspace{-1mm}
 \label{Fig4}
\end{figure}

\begin{figure}[htbp]

\subfigure
{

\begin{minipage}[h]{0.45\linewidth}
 \includegraphics[width=1.65in]{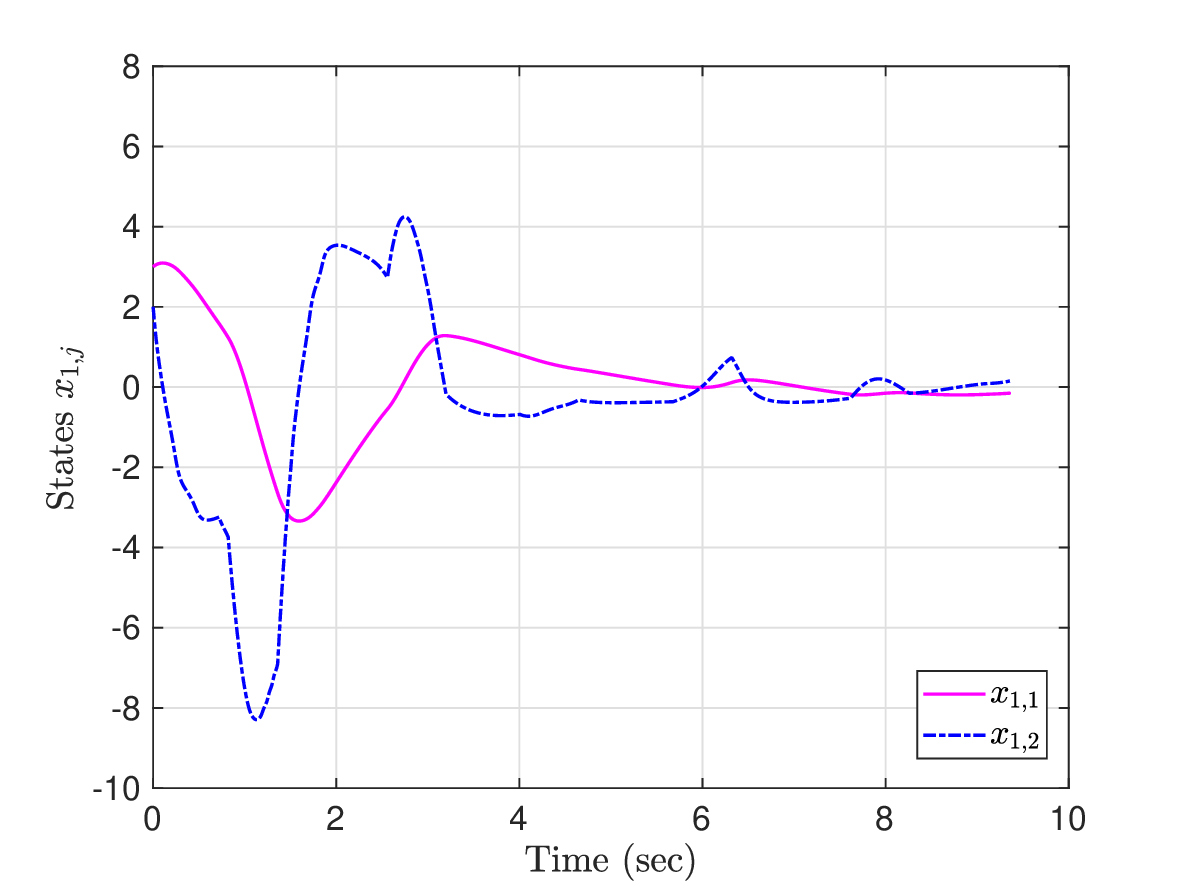}
 \includegraphics[width=1.65in]{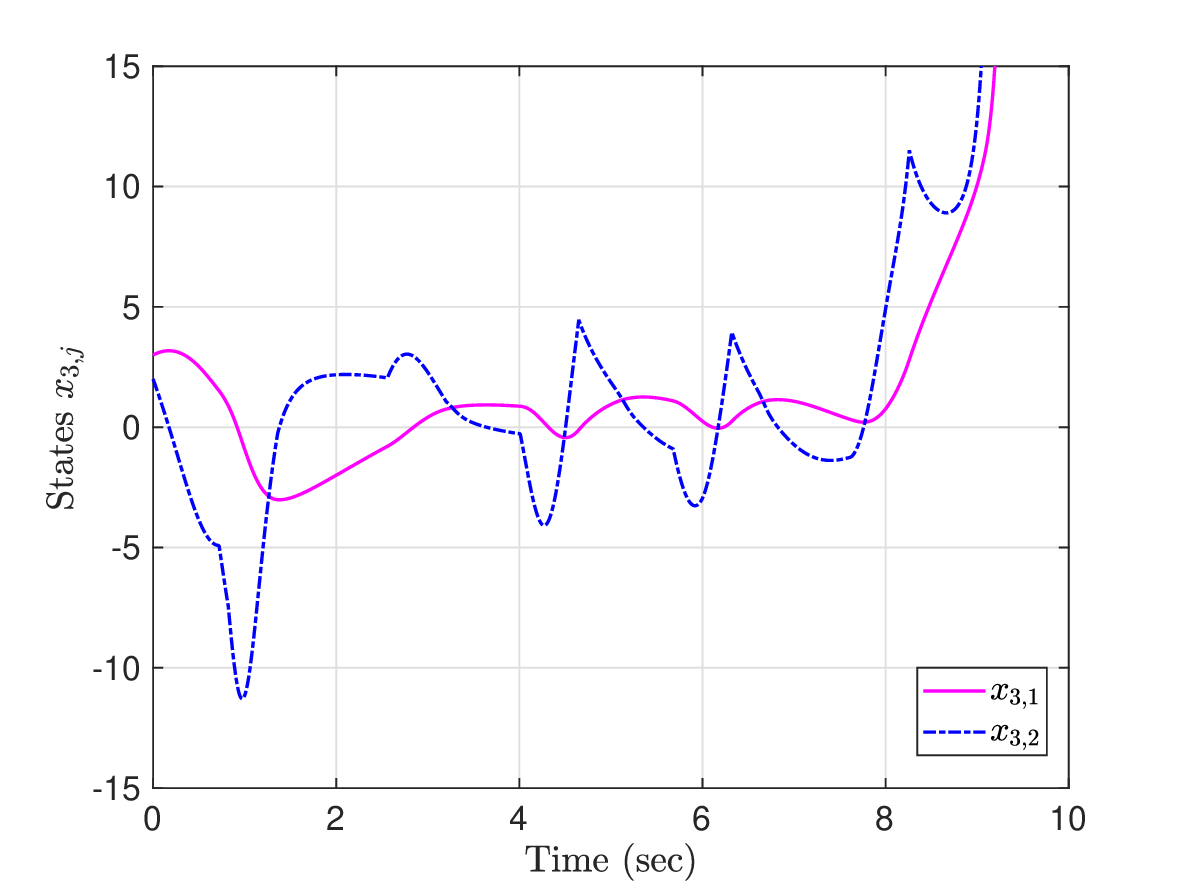}
 \end{minipage}
 \begin{minipage}[h]{0.45\linewidth}
 \includegraphics[width=1.65in]{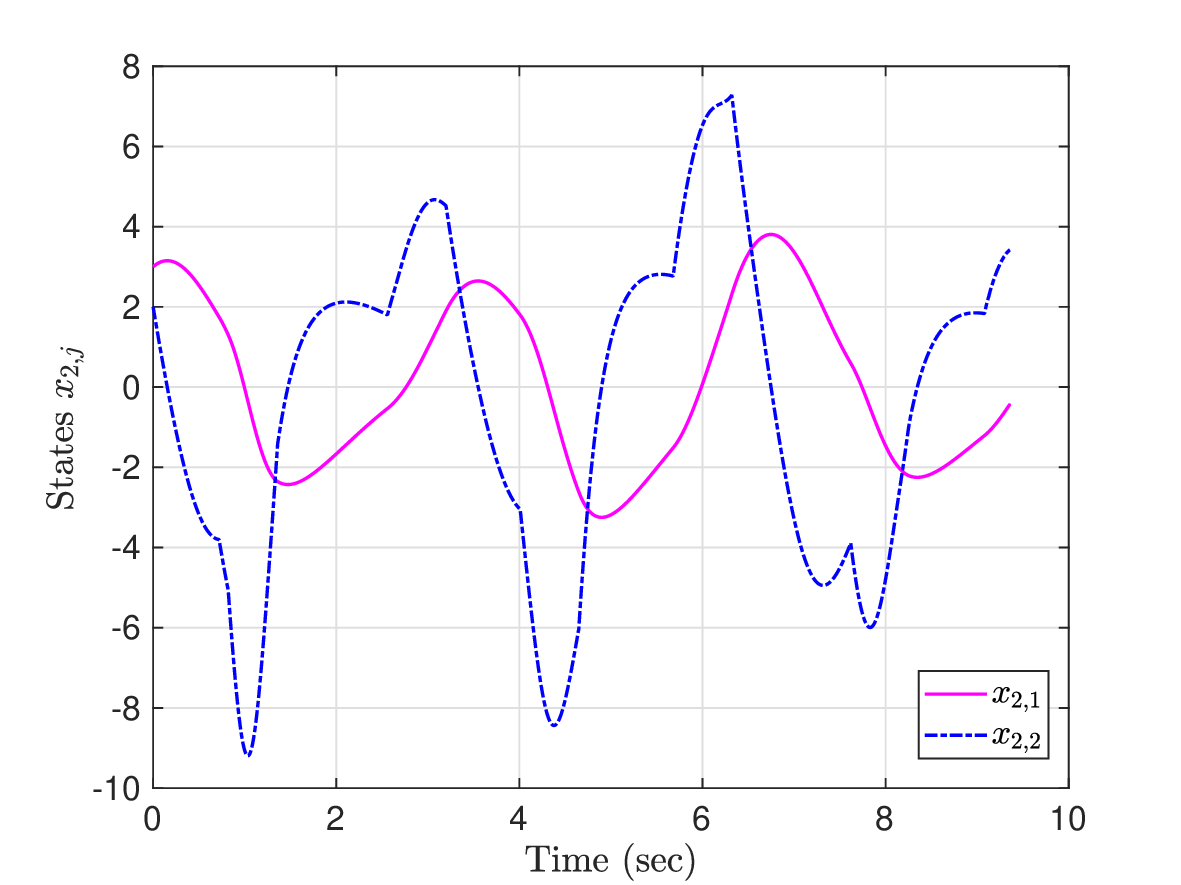}
  \includegraphics[width=1.65in]{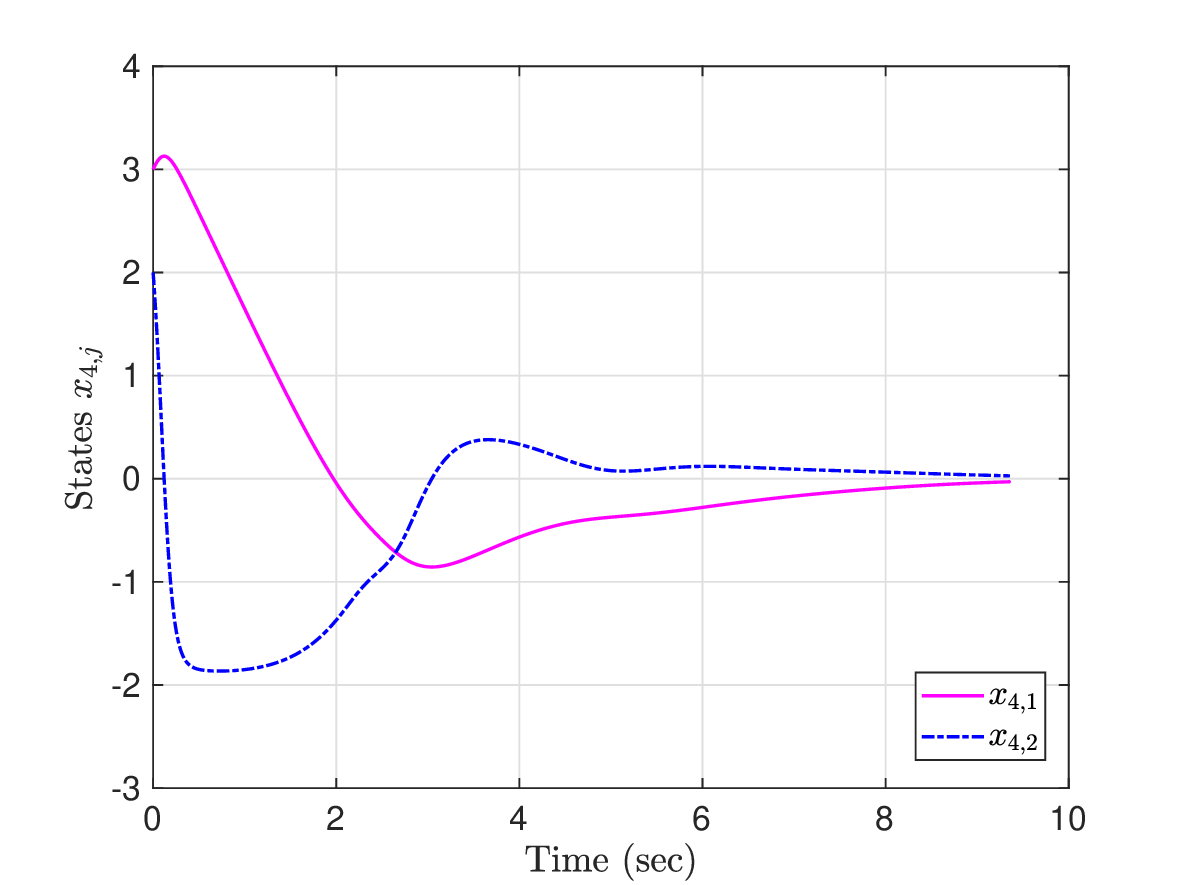}
\end{minipage}
}
 \caption{Trajectory of system states $ x_{i,j}$ with long anomaly time}
\label{Fig5}
\end{figure}

Let the $i={1,2,3}$ subsystems to suffer the network anomalies, for the time duration of each anomaly, we set the duration of each anomaly as $ t_{d,h}=0.64s $. Six anomalies begin at $ t_1=0.72s $, $ t_2=2.56s $, $ t_3=4.01s $, $ t_4=5.68s $, $ t_5=7.62s $, and $ t_6=9.08s $ respectively, as shown in~Fig. \ref{Fig4}. We observe that the subsystem states $x_{i,j}\left (i=1,2,3,4;j=1,2\right )$ in~Fig. \ref{Fig5} cannot be completely maintained stable due to the long time duration of each anomaly.

\begin{figure}[htbp]
 \centering
 \includegraphics[width=3.3in]{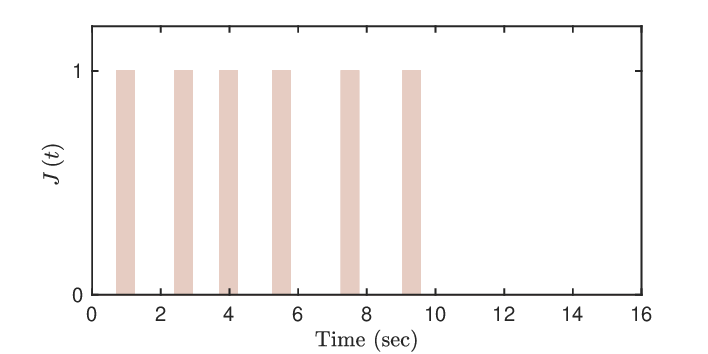}
 \caption{Bounded anomaly time intervals.}
 \vspace{-1mm}
 \label{Fig6}
\end{figure}

\begin{figure}[htbp]

\subfigure
{\begin{minipage}[h]{0.45\linewidth}
 \includegraphics[width=1.65in]{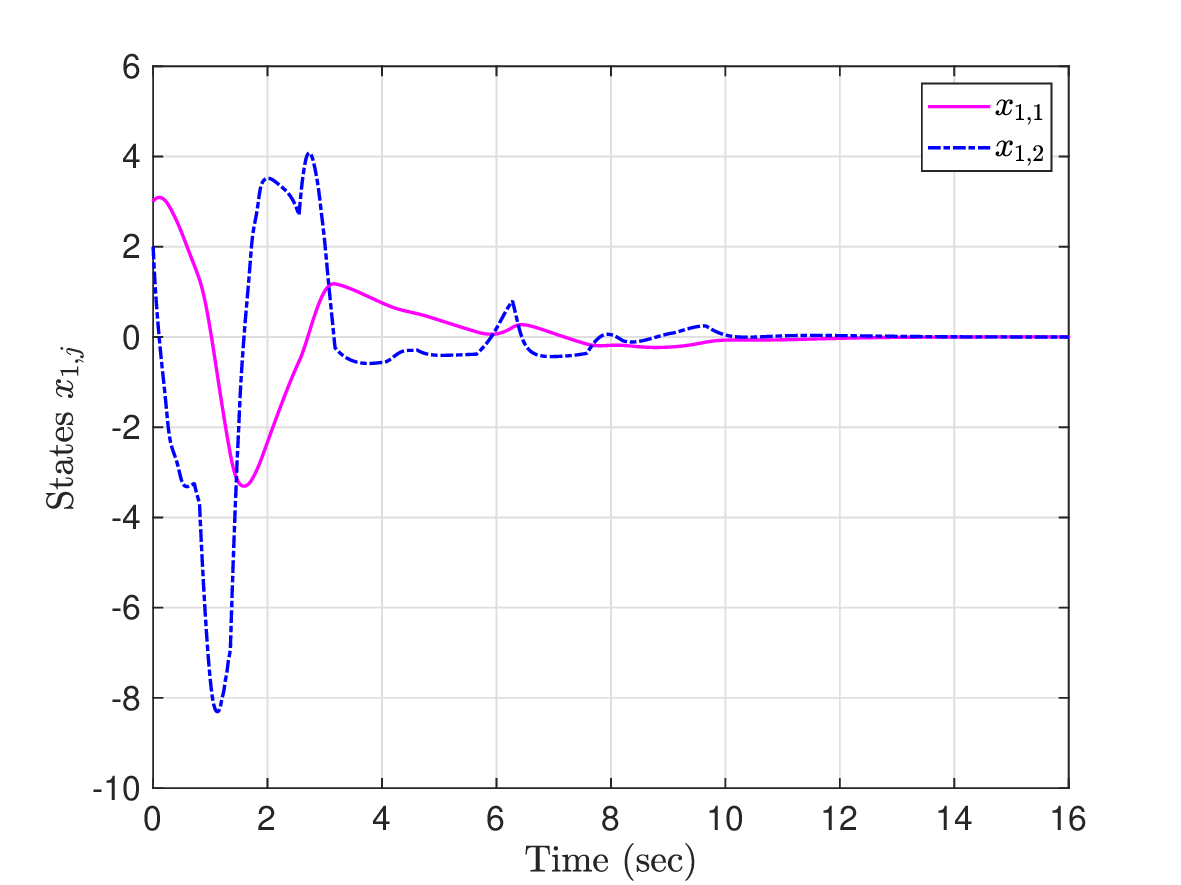}
 \includegraphics[width=1.65in]{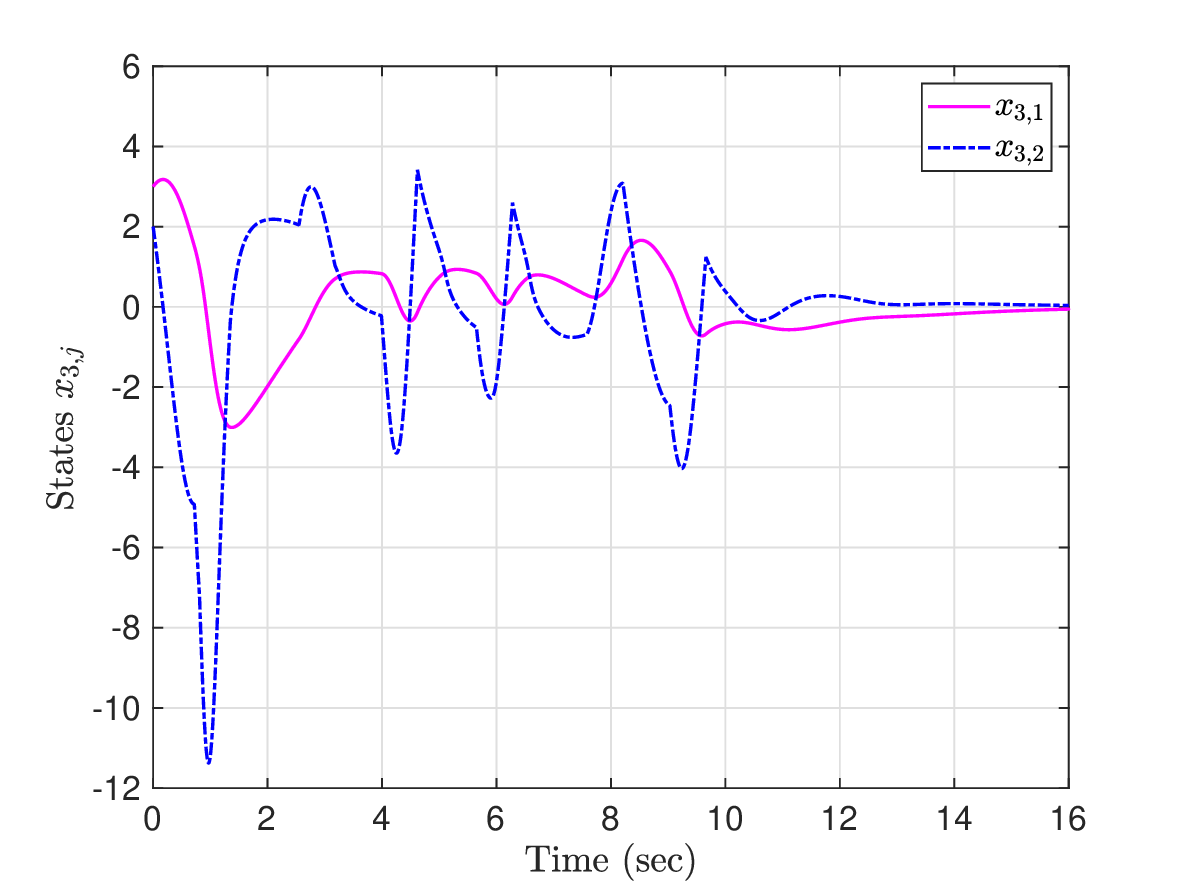}
 \end{minipage}
 \begin{minipage}[h]{0.45\linewidth}
 \includegraphics[width=1.65in]{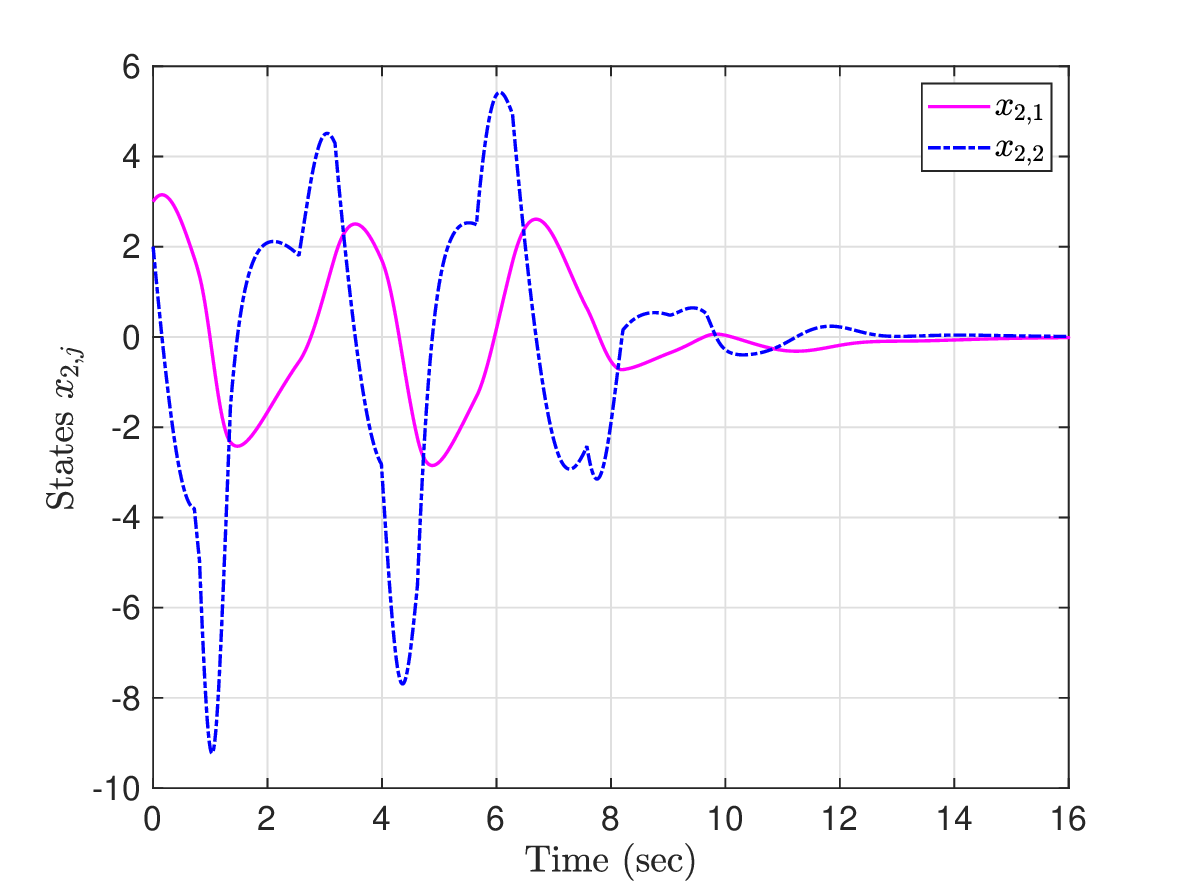}
  \includegraphics[width=1.65in]{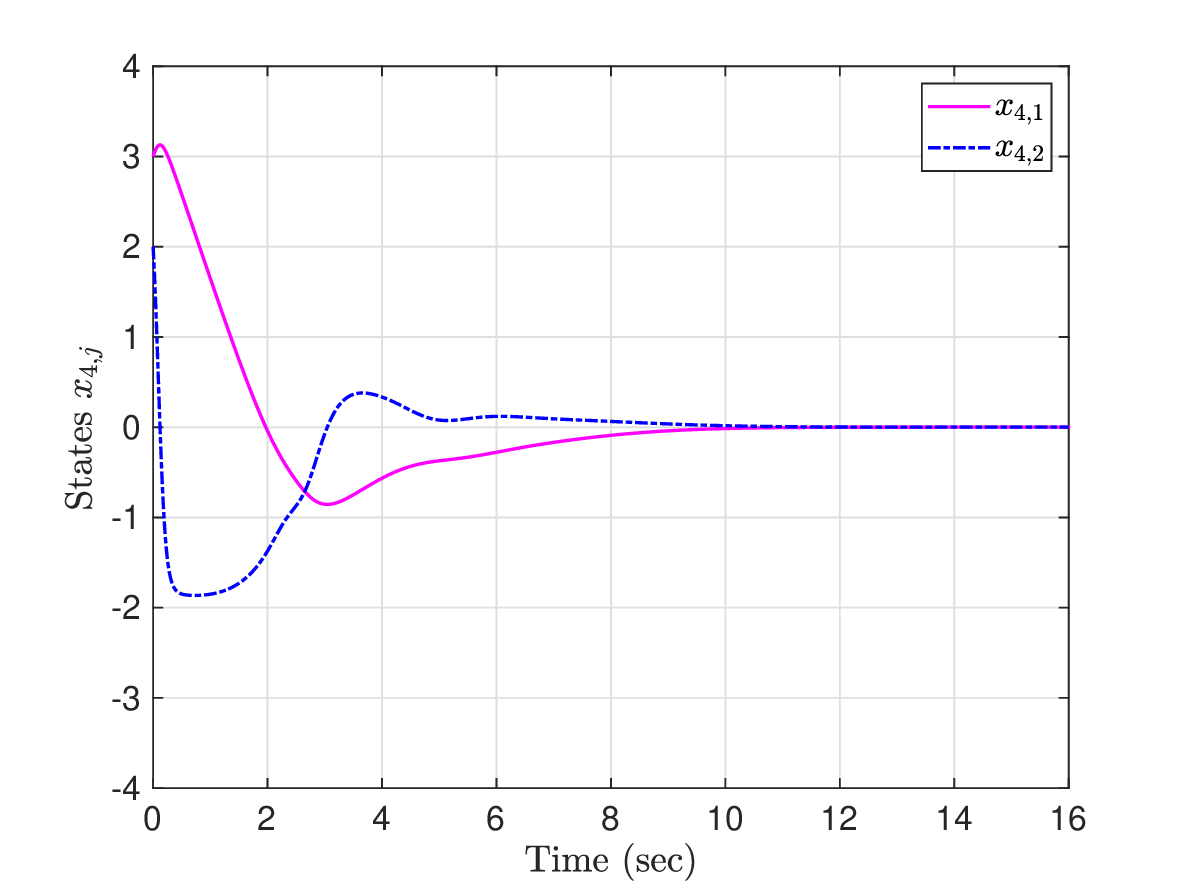}
\end{minipage}}
 \caption{Trajectory of system states $ x_{i,j}$ with bounded anomaly duration}
\label{Fig7}
\end{figure}

Instead of the above, we set the bounded duration of each anomaly as $ t_{d,h}=0.63s $. Six anomalies begin at $ t_1=0.72s $, $ t_2=2.55s $, $ t_3=3.99s $, $ t_4=5.65s $, $ t_5=7.58s $, and $ t_6=9.03s $ respectively, as shown in~Fig. \ref{Fig6}. The convergent trajectories of the subsystem states $x_{i,j}\left (i=1,2,3,4;j=1,2\right )$ as shown in~Fig. \ref{Fig7}. Observe that the states converge asymptotically to zero. Based on the data presented in~Table \ref{table1}, we observe that the mean squared error (MSE) associated with bounded anomaly duration is smaller than that observed with long anomaly duration. It shows that, given the controller design and stability analysis of the interconnected system, if each anomaly time is less than a certain conditions, and regardless of whether matter any subsystems of the interconnected system are anomalous, all signals of the interconnected system remain asymptotically stable.

\begin{table}[htbp]
    \centering
      \caption{Anomaly time MSE}
    \label{table1}
    \begin{tabular}{l c c}
        \toprule
         \ &Long anomaly time & Bounded anomaly time  \\
        \midrule
        $x_{11}$   & 1.61   & 0.92 \\
        $x_{21}$   & 4.36   & 1.77 \\
        $x_{31}$ &  $2.07\times10^2$    & 1.13 \\
        $x_{41}$ &  0.93  & 0.54 \\

        \bottomrule
    \end{tabular}

\end{table}

\subsection{Simulation of the resting time between two consecutive anomalies}
\begin{figure}[ht]
 \centering
 \includegraphics[width=3.3in]{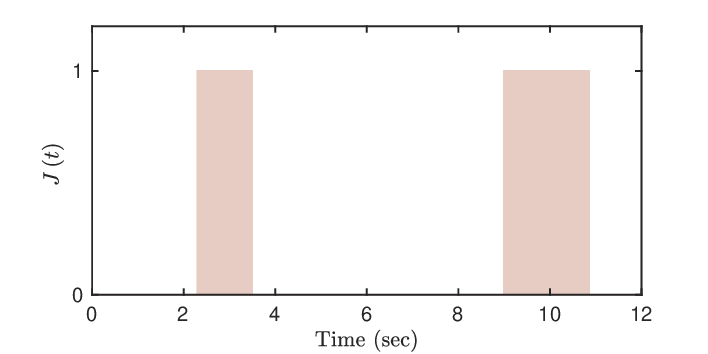}
 \caption{Two anomalies with insufficient anomaly resting time}
 \vspace{-1mm}
 \label{Fig8}
\end{figure}

\begin{figure}[htbp]
\subfigure
{

\begin{minipage}[h]{0.45\linewidth}
 \includegraphics[width=1.65in]{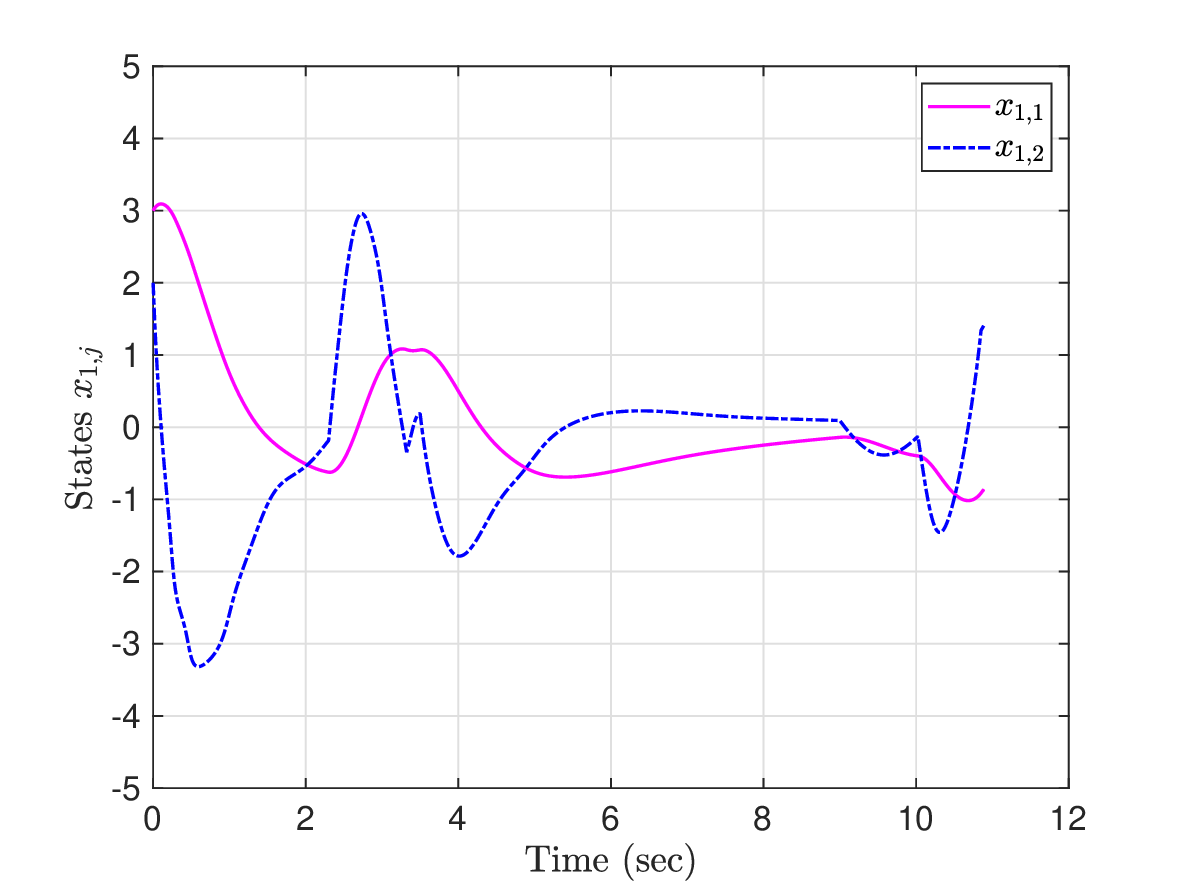}
 \includegraphics[width=1.65in]{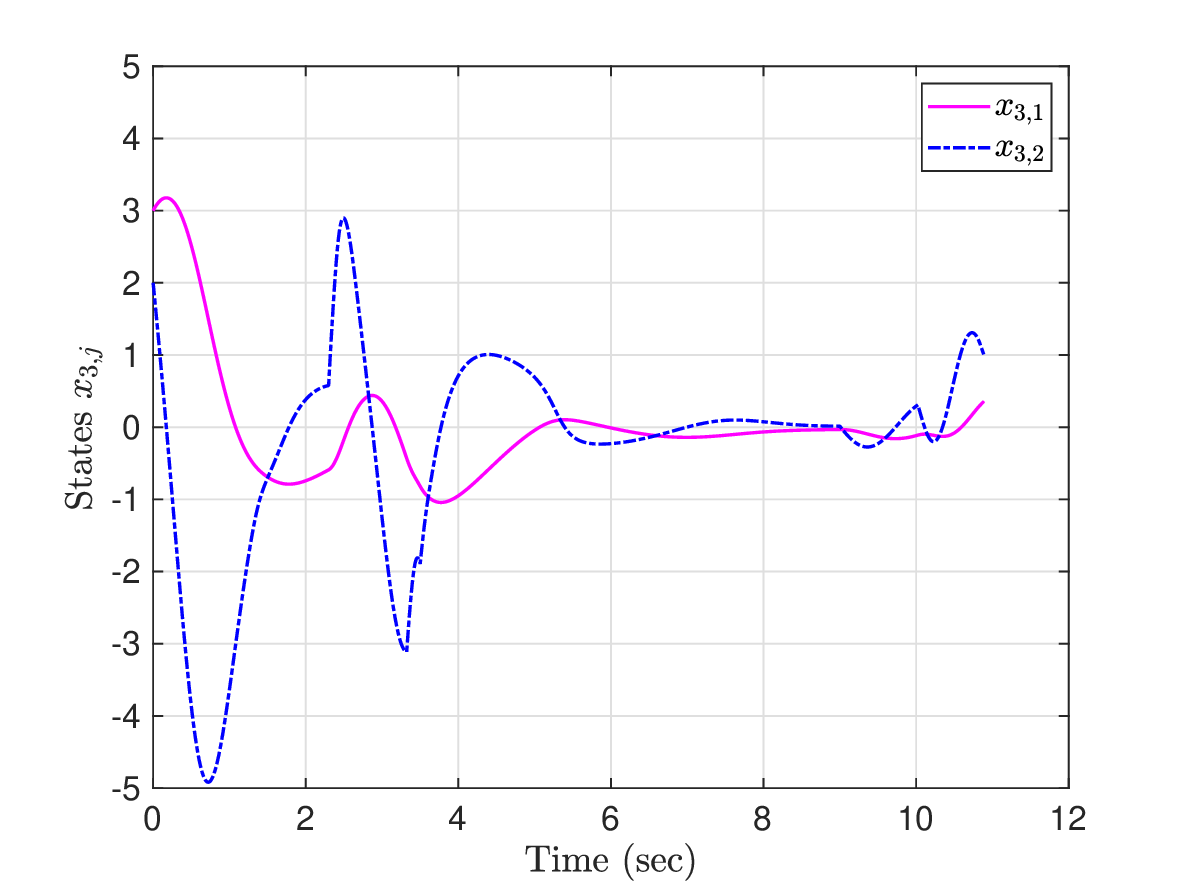}
 \end{minipage}
 \begin{minipage}[h]{0.45\linewidth}
 \includegraphics[width=1.65in]{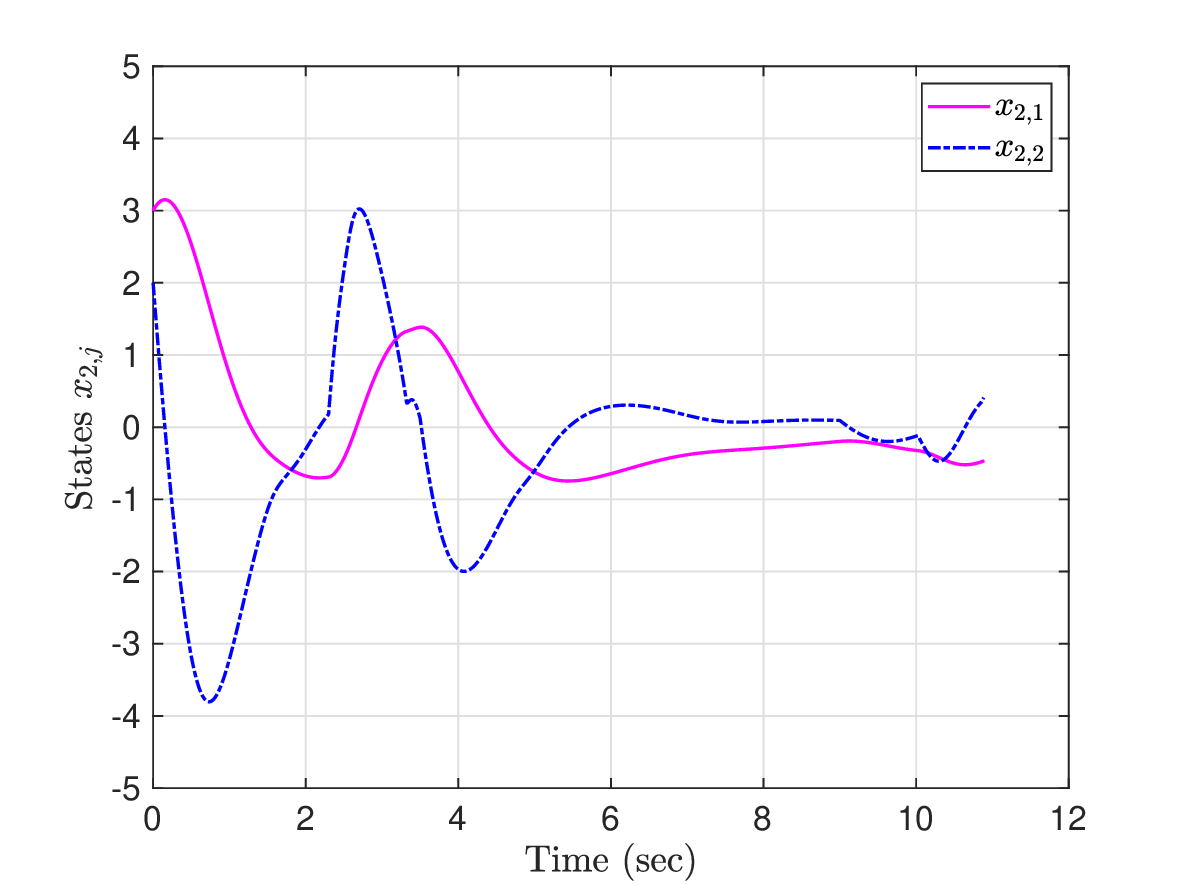}
  \includegraphics[width=1.65in]{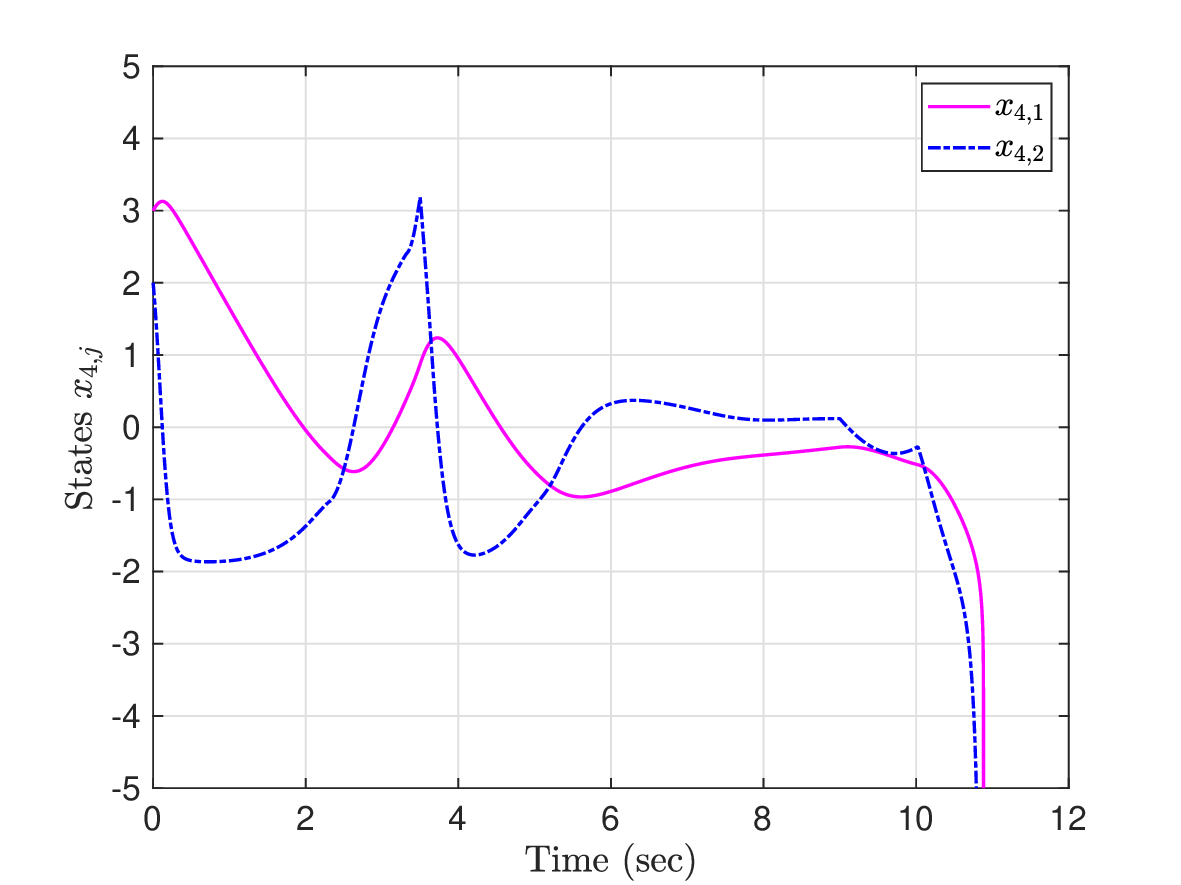}
\end{minipage}
}
 \caption{Trajectory of system states $ x_{i,j}$ with insufficient anomaly resting time}
\label{Fig9}
\end{figure}

Let the $i={1,2,3,4}$ subsystems to suffer the network anomalies, for the resting time between two consecutive anomalies, we launch two network anomalies as shown in~Fig. \ref{Fig8}. The first anomaly begins at $ t_1=2.3s $ and $ t_{d,1}=1.2s $, after the first anomaly is over with the resting time $t_{r,1}=5.5s $, followed by a second anomaly. The second anomaly begins at $ t_2=9.0s $ and $ t_{d,2}=1.85s $. We observe that the subsystem states $x_{i,j}\left (i=1,2,3,4;j=1,2\right )$ in~Fig. \ref{Fig9} cannot be completely maintained within the safety set due to the insufficient resting time between two consecutive anomalies.


\begin{figure}[htbp]
 \centering
 \includegraphics[width=3.3in]{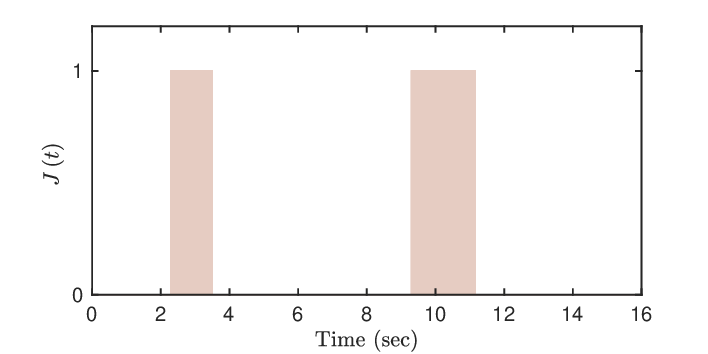}
 \caption{Two anomalies with sufficient anomaly resting time}
 \vspace{-1mm}
 \label{Fig10}
\end{figure}

\begin{figure}[htbp]

\subfigure
{\begin{minipage}[h]{0.45\linewidth}
 \includegraphics[width=1.65in]{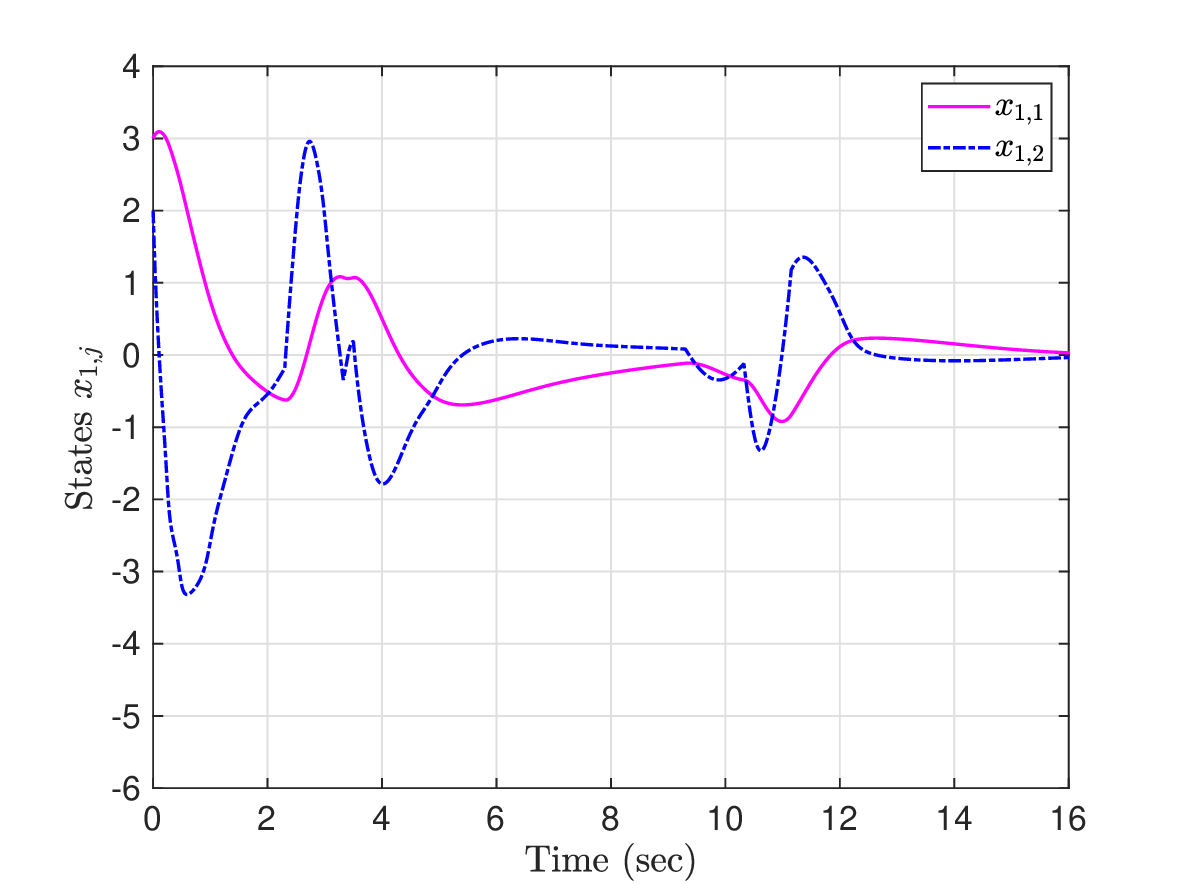}
 \includegraphics[width=1.65in]{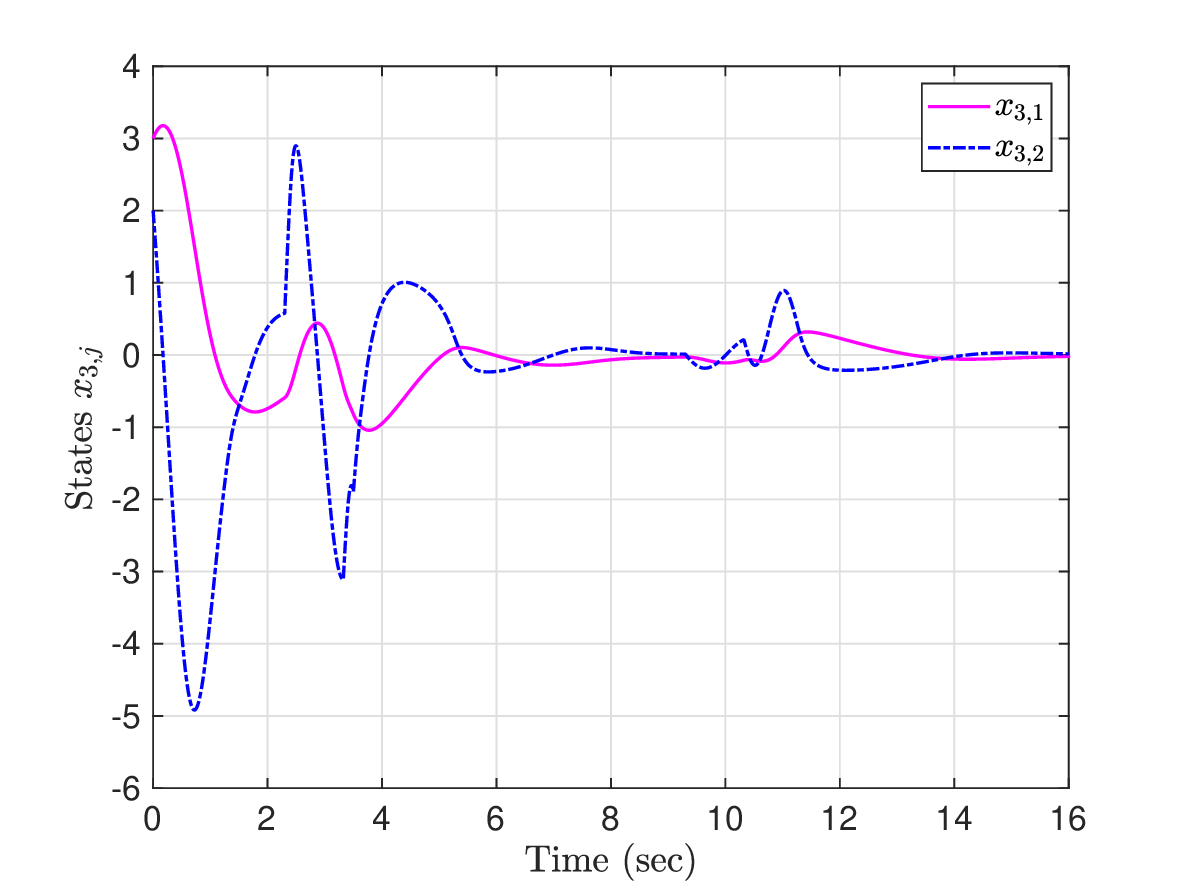}
 \end{minipage}
 \begin{minipage}[h]{0.45\linewidth}
 \includegraphics[width=1.65in]{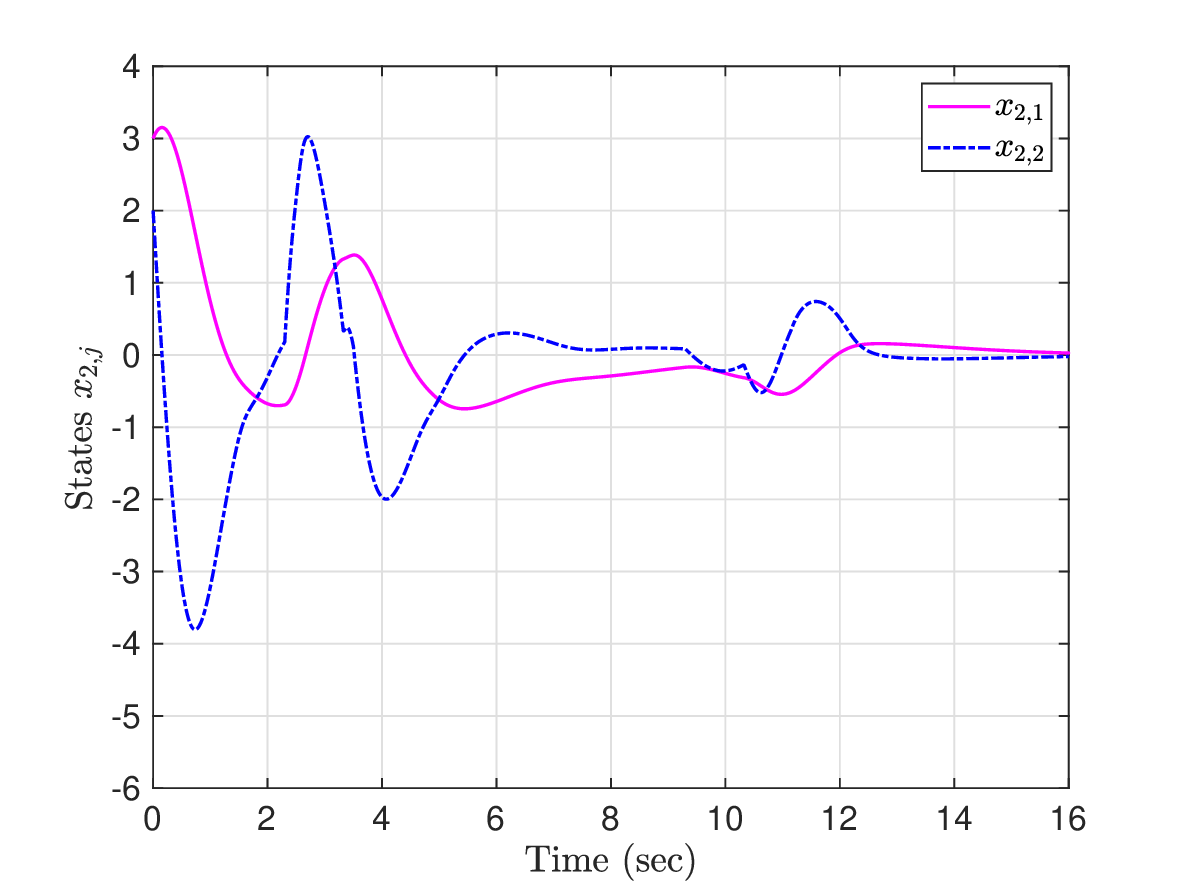}
  \includegraphics[width=1.65in]{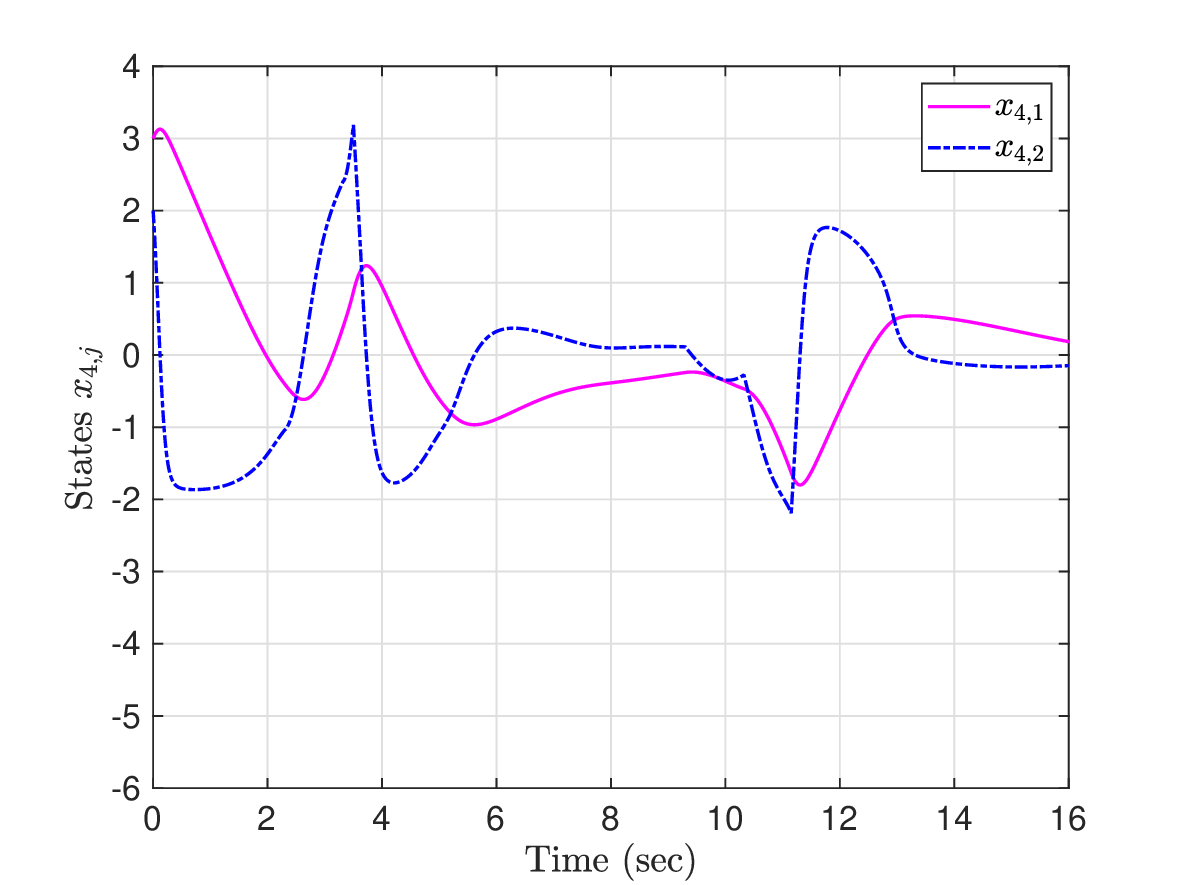}
\end{minipage}}
 \caption{Trajectory of system states $ x_{i,j}$ with sufficient anomaly resting time}
\label{Fig11}
\end{figure}

Keeping the same network anomaly duration, we give sufficient resting time as shown in~Fig. \ref{Fig10}. The first anomaly begins at $ t_1=2.3s $ and $ t_{d,1}=1.2s $, after the first anomaly is over with the resting time $t_{r,1}=5.8s $. The second anomaly begins at $ t_2=9.3s $ and $ t_{d,2}=1.85s $. The subsystem states $x_{i,j}\left (i=1,2,3,4;j=1,2\right )$ are shown in~Fig. \ref{Fig11}, which demonstrates that as long as the resting time is sufficient, all signals can remain bounded under consecutive network anomalies. Based on the data in~Table \ref{table2}, we can conclude that the MSE associated with sufficient resting time is smaller than with insufficient resting time. It shows that, given the controller design and stability analysis of the interconnected system, if the resting time between consecutive anomalies meets certain conditions, and regardless of whether matter any subsystems of the interconnected system are anomalous, all signals of the interconnected system remain bounded.

\begin{table}[htbp]
    \centering
      \caption{Resting time MSE}
    \label{table2}
    \begin{tabular}{l c c}
        \toprule
         \ &Insufficient resting time & Sufficient resting time \\
        \midrule
        $x_{11}$   & 0.76   & 0.52 \\
        $x_{21}$   & 0.85   & 0.58 \\
        $x_{31}$ &  0.67    & 0.46 \\
        $x_{41}$ &  $7.31\times10^{164}$  & 0.85 \\

        \bottomrule
    \end{tabular}

\end{table}

\section{Conclusions}\label{s6}
We investigate a class of nonlinear interconnected systems consisting of uncertain dynamical subsystems possibly subject to multiple network anomalies. In the framework of adaptive approximation, a back-stepping approach is used to design the distributed controller and the parameter estimator
The stability of the anomalous and normal subsystems is analyzed, where both local network anomalies and neighboring anomaly propagation effects are considered. In order to quantify the resilience of the interconnected systems under network anomalies, we derive the bounds to the duration of each anomaly and to the resting time between two consecutive anomalies. Specifically, when the duration of each anomaly is less than the boundary constant of the resilience condition, the interconnected systems based on the distributed approximation controller remains asymptotically stable. Alternatively, if the resting time between two consecutive anomalies is larger than the boundary constant associated with the resilience condition, then all signals of the interconnected systems are guaranteed to be bounded. These results show that under the action of the proposed distributed adaptive control scheme, the interconnected system can maintain stable under both qualitative and quantitative resilience conditions in the presence of network anomalies. Our simulation results verify the validity of the theoretical analysis.

Future research will focus on analyzing the system resilience condition under different classes of network anomalies and the active performance enhancement against network anomalies for multi-agent systems.


\section*{References}
\vspace{-1em}
\begin{thebibliography}{00}\leftskip1pc
\bibitem{s1}
G. Wen, Y. Lv, W. X. Zheng, J. Zhou and J. Fu, ``Joint robustness of time-varying networks and its applications to resilient consensus''. \emph{IEEE Transactions on Automatic Control}, vol. 68, no. 11, pp. 6466-6480, 2023.
\bibitem{a1}
M. Zhu and S. Mart{\'\i}nez, ``On distributed constrained formation control in operator--vehicle
  adversarial networks''. \emph{Automatica}, vol. 49, no. 12, pp. 3571-3582, 2013.

\bibitem{a2}
Y. Yu, W. Yang, W. Ding and J. Zhou, ``Reinforcement learning solution for cyber-physical systems security
against replay attacks''. \emph{IEEE Transactions on Information Forensics and Security}, vol. 18, pp. 2583-2595, 2023.

\bibitem{a3}
C. Fang, Y. Qi, P. Cheng and W. X. Zheng, ``Optimal periodic watermarking schedule for replay attack detection in cyber-physical systems''. \emph{Automatica}, vol. 112, 108698, 2020.

\bibitem{a4}

S. X. Ding, L. Li, D. Zhao, C. Louen and T. Liu, ``Application of the unified control and detection framework to detecting stealthy integrity cyber-attacks on feedback control systems''. \emph{Automatica}, vol. 142, 110352, 2022.


\bibitem{a5}
T. Li, Z. Wang, L. Zou, B. Chen and L. Yu , ``A dynamic encryption-decryption scheme for replay attack detection in cyber-physical systems''. \emph{Automatica}, vol. 151, 110926, 2023.

\bibitem{a6}

A. Barboni, H. Rezaee, F. Boem, and T. Parisini, ``Detection of covert cyber-attacks in interconnected systems: A distributed model-based approach''. \emph{IEEE Transactions on Automatic Control}, vol. 65, no. 9, pp. 3728-3741, 2020.

\bibitem{a7}
A. Naha, A. Teixeira, A. Ahl{\'e}n and S. De, ``Sequential detection of replay attacks''. \emph{IEEE Transactions on Automatic Control}, vol. 68, no. 3, pp. 1941-1948, 2023.

\bibitem{a8}

M. Zhu and S. Martinez, ``On the performance analysis of resilient networked control systems
  under replay attacks''. \emph{IEEE Transactions on Automatic Control}, vol. 59, no. 3, pp. 804-808, 2014.

\bibitem{a9}

X. Xu, X. Li, P. Dong, Y. Liu and H. Zhang, ``Robust reset speed synchronization control for an integrated
  motor-transmission powertrain system of a connected vehicle under a replay
  attack''. \emph{IEEE Transactions on Vehicular Technology}, vol. 70, no. 6, pp. 5524-5536, 2021.

\bibitem{a10}

Q. Zhong, J. Yang, K. Shi and S. Zhong, ``Design of observer-based discrete-type PID control for reconstructed jump model of interconnected power system under hybrid attacks''. \emph{IEEE Transactions on Smart Grid}, vol. 14, no. 3, pp. 1896-1906, 2023.

\bibitem{a11}

W. Yao, Y. Wang, Y. Xu and C. Deng, ``Cyber-resilient control of an islanded microgrid under latency attacks and random DoS attacks''. \emph{IEEE Transactions on Industrial Informatics}, vol. 19, no. 4, pp. 5858-5869, 2023.

\bibitem{a12}

L. Dong, H. Xu, X. Wei, X. Hu, ``Security correction control of stochastic cyber-physical systems
  subject to false data injection attacks with heterogeneous effects''. \emph{ISA Transactions}, vol. 123, pp. 1-13, 2022.

\bibitem{a13}

Y. Yang, J. Huang, X. Su and B. Deng, ``Adaptive control of cyber-physical systems under deception and
  injection attacks''. \emph{Journal of the Franklin Institute}, vol. 358, no. 12, pp. 6174-6194, 2021.

\bibitem{a14}

F. Farivar, M. S. Haghighi, A. Jolfaei and M. Alazab, ``Artificial intelligence for detection, estimation, and compensation of malicious attacks in nonlinear cyber-physical systems and industrial IoT''. \emph{IEEE Transactions on Industrial Informatics}, vol. 16, no. 4, pp. 2716-2725, 2020.

\bibitem{a15}

X. M. Zhang, Q. L. Han, X. Ge and L. Ding, ``Resilient control design based on a sampled-data model for a class of networked control systems under denial-of-service attacks''. \emph{IEEE Transactions on Cybernetics}, vol. 50, no. 8, pp. 3616-3626, 2020.
\bibitem{a16}

X. Jin, S. L{\"u}, J. Qin, W. X. Zheng and Q. Liu, ``Adaptive ELM-based security control for a class of
  nonlinear-interconnected systems with DoS attacks''. \emph{IEEE Transactions on Cybernetics}, vol. 53, no. 8, pp. 5000-5012, 2023.
\bibitem{a17}

L. An and G. H. Yang, ``Decentralized adaptive fuzzy secure control for nonlinear uncertain
  interconnected systems against intermittent DoS attacks''. \emph{IEEE Transactions on Cybernetics}, vol. 49, no. 3, pp. 827-838, 2019.
\bibitem{a18}

Y. Tan, Q. Liu, D. Du, B. Niu and S. Fei, ``Observer-based finite-time $H_\infty$ control for interconnected fuzzy
  systems with quantization and random network attacks''. \emph{IEEE Transactions on Fuzzy Systems}, vol. 29, no. 3, pp. 674-685, 2021.
\bibitem{s2}
D. Zhao, Y. Shi, Y. Li and S. Liu, ``Fault accommodation of multiple faults for a class of nonlinear uncertain systems: a dynamic fault isolation information framework''. \emph{IEEE Transactions on Automatic Control}, pp. 1-8, 2024.
\bibitem{a19}

X. Shao and D. Ye , ``Neural-network-based adaptive secure control for nonstrict-feedback
  nonlinear interconnected systems under DoS attacks''. \emph{Neurocomputing}, vol. 448, pp. 263-275, 2021.

\bibitem{a20}

X. Jin, W. M. Haddad and T. Yucelen, ``An adaptive control architecture for mitigating sensor and actuator attacks in cyber-physical systems''. \emph{IEEE Transactions on Automatic Control}, vol. 62, no. 11, pp. 6058-6064, 2017.

\bibitem{a21}

X. M. Li, W. Xiao, G. Lin, H. Li and R. Lu, ``Observer-based security control for distributed cyber-physical systems under replay attacks''. \emph{International Journal of Robust and Nonlinear Control}, vol. 33, no. 13, pp. 8015-8035, 2023.
\bibitem{a22}

W. Xu, J. Kurths, G. Wen and X. Yu, ``Resilient event-triggered control strategies for second-order
  consensus''. \emph{IEEE Transactions on Automatic Control}, vol. 67, no. 8, pp. 4226-4233, 2022.


\bibitem{a23}

T. Yin, Z. Gu and X. Xie, ``Observer-based event-triggered sliding mode control for secure formation tracking of multi-UAV systems''. \emph{IEEE Transactions on Network Science and Engineering}, vol. 10, no. 2, pp. 887-898, 2023.
\bibitem{a24}

L. Su, D. Ye and X. Zhao, ``Static output feedback secure control for cyber-physical systems
  based on multisensor scheme against replay attacks''. \emph{International Journal of Robust and Nonlinear Control}, vol. 30, no. 18, pp. 8313-8326, 2020.
\bibitem{a25}

J. Huang, L. Zhao and Q. G. Wang, ``Adaptive control of a class of strict feedback nonlinear systems
  under replay attacks''. \emph{ISA Transactions}, vol. 107, pp. 134-142, 2020.

\bibitem{a26}
F. Zhu and Z. Han, ``A note on observers for Lipschitz nonlinear systems''. \emph{IEEE Transactions on Automatic Control}, vol. 47, no. 10, pp. 1751-1754, 2002.

\bibitem{a27}
J. A. Farrell and M. M. Polycarpou, \emph{Adaptive Approximation based Control: Unifying Neural, Fuzzy and Traditional Adaptive Approximation Approaches}. New York, NY, USA: John Wiley \& Sons, 2006.

\bibitem{a28}

M. Krstic, I. Kanellakopoulos, and P. V. Kokotovic, \emph{Nonlinear and Adaptive Control Design}. New York, NY, USA: Wiley, 1995.
\bibitem{s3}

W. Wang, J. Long, J. Zhou, J. Huang and C. Wen, ``Adaptive backstepping based consensus tracking of uncertain nonlinear systems with event-triggered communication''. \emph{Automatica}, vol. 133, 109841, 2021.
\bibitem{a29}

P. Tabuada, ``Event-triggered real-time scheduling of stabilizing control tasks''. \emph{IEEE Transactions on Automatic Control}, vol. 52, no. 9, pp. 1680-1685, 2007.
\bibitem{a30}

L. Cui, Y. Zhang, X. Wang and X. Xie, ``Event-triggered distributed self-learning robust tracking control for
  uncertain nonlinear interconnected systems''. \emph{Applied Mathematics and Computation}, vol. 395, no. 15, 125871, 2021.
\bibitem{a31}

M. M. Polycarpou, ``Fault accommodation of a class of multivariable nonlinear dynamical
  systems using a learning approach''. \emph{IEEE Transactions on Automatic Control}, vol. 46, no. 5, pp. 736-742, 2001.

\bibitem{a32}

D. Zhao and M. M. Polycarpou, ``Distributed fault accommodation of multiple sensor faults for a class
  of nonlinear interconnected systems''. \emph{IEEE Transactions on Automatic Control}, vol. 67, no. 4, pp. 2092-2099, 2022.


\end{thebibliography}
\end{document}